\documentclass[a4paper,aps,prd,twocolumn,eqsecnum,amsmath,amssymb,nofootinbib,superscriptaddress]{revtex4-2}
\usepackage{dcolumn}
\usepackage{bm}
\usepackage{graphics}
\usepackage{graphicx}
\graphicspath{{figures/}}
\usepackage{epsfig}
\usepackage{booktabs,multirow}
\usepackage{hhline}
\usepackage{slashed}
\usepackage{rccol}
\usepackage{mathrsfs}
\usepackage{amsmath, amssymb}
\usepackage[dvipsnames]{xcolor,colortbl}
\usepackage{subfigure}
\newcommand{\orcid}[1]{\href{https://orcid.org/#1}{\includegraphics[width=10pt]{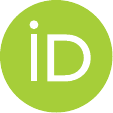}}}

\usepackage{xcolor}\colorlet{linkequation}{blue}
\usepackage[colorlinks=true, 
linkcolor=Red,   
filecolor=Red,   
citecolor=Green, 
urlcolor=Blue,
hyperfootnotes=true]{hyperref}

\newcommand*{\SavedEqref}{}
\let\SavedEqref\eqref
\renewcommand*{\eqref}[1]{%
	\begingroup
	\hypersetup{
		linkcolor=linkequation,
		linkbordercolor=linkequation,
	}%
	\SavedEqref{#1}%
	\endgroup
}

\begin{document}
\title{Probing Physics Beyond the Standard Model with Precision Reactor CE$\nu$NS Data from CONUS+}

\author{Mehmet~Demirci \orcid{0000-0003-2504-6251}}
\email{mehmetdemirci@ktu.edu.tr } 
\affiliation{Department of Physics, Karadeniz Technical University, Trabzon 61080, Türkiye}

\date{\today}
\begin{abstract}
The CONUS+ experiment has recently reported a precision measurement of coherent elastic neutrino-nucleus scattering (CE$\nu$NS) using low-threshold, high-purity germanium detectors exposed to reactor electron antineutrinos. Based on the Run~2 dataset with a combined reactor-on exposure of $1296~\mathrm{kg\cdot d}$ across three detectors (C3, C6, and C7), CONUS+ measured the CE$\nu$NS signal with a relative precision of $9.9\%$, consistent with the Standard Model (SM) expectation. In this work, we perform a simultaneous binned Poisson likelihood analysis of the CONUS+ Run~2 spectra to probe a broad range of physics scenarios beyond the SM. Specifically, we derive updated constraints on light vector ($U(1)_{B-L}$, $U(1)_{B-2L_e-L_\mu}$, $U(1)_{B-3L_e}$, and universal $Z'$) and scalar mediators, neutrino electromagnetic properties---namely the effective magnetic moment, charge radius, and electric millicharge---as well as flavor-conserving and flavor-changing vector nonstandard neutrino interactions (NSIs). These results substantially improve upon previous reactor and spallation-source CE$\nu$NS bounds across most of the probed parameter space, highlighting the growing sensitivity of sub-keV-threshold germanium detectors to new neutrino interactions.
\end{abstract}

\maketitle

\section{Introduction}
\label{sec:intro}

Coherent elastic neutrino-nucleus scattering (CE$\nu$NS) is a Standard Model (SM) neutral-current process in which a low-energy neutrino interacts coherently with an atomic nucleus as a single entity~\cite{Freedman:1973yd,Drukier:1984vhf}. Although the coherent enhancement produces a comparatively large cross section scaling approximately with the square of the neutron number, experimental detection remained elusive for decades because the coherence condition restricts the resulting nuclear recoil energies to the sub-keV to few-keV scale. The first observation of CE$\nu$NS was achieved by the COHERENT Collaboration using stopped-pion neutrinos at a spallation neutron source with a CsI[Na] scintillator~\cite{COHERENT:2017ipa}, followed by subsequent detections on liquid argon (LAr)~\cite{COHERENT:2020iec} and germanium~\cite{COHERENT:2024axu} targets. More recently, the experimental landscape has expanded rapidly to nuclear reactors with measurements from Dresden-II~\cite{Colaresi:2022obx} and CONUS+~\cite{Ackermann:2025obx}, as well as to liquid-xenon dark matter direct detection facilities observing solar $^8$B neutrino-induced CE$\nu$NS at XENONnT~\cite{XENON:2024ijk}, PandaX-4T~\cite{PandaX:2024muv}, and LZ~\cite{LZ:2025igz}. Collectively, these breakthroughs have opened a vibrant phenomenological program to test electroweak interactions and fundamental properties of neutrinos within the SM, as well as to search for physics beyond the SM (BSM).

Commercial nuclear reactors provide an intense, pure flux of electron antineutrinos ($\bar{\nu}_e$) in the MeV energy range, primarily generated by the $\beta$-decays of neutron-rich fission fragments of $^{235}\mathrm{U}$, $^{239}\mathrm{Pu}$, $^{238}\mathrm{U}$, and $^{241}\mathrm{Pu}$, along with an additional low-energy component from neutron captures below the inverse beta decay threshold. Because reactor antineutrinos carry significantly lower energies than spallation-source neutrinos, the coherence condition is almost fully preserved across the entire recoil spectrum. Consequently, uncertainties associated with the finite nuclear size and nuclear form factors are drastically suppressed, making low-threshold reactor detectors particularly sensitive to spectral distortions induced by BSM physics at small momentum transfers. Alongside the first reactor observations, a broad suite of dedicated experiments—including CONNIE~\cite{CONNIE:2019swq}, NUCLEUS~\cite{NUCLEUS:2019igx}, $\nu$GEN~\cite{nGeN:2022uje}, NEON~\cite{NEON:2022hbk}, Ricochet~\cite{Ricochet:2023nvt}, TEXONO~\cite{TEXONO:2024vfk}, RED-100~\cite{RED-100:2024izi}, and MINER~\cite{MINER:2025wrk}—continue to advance reactor-based CE$\nu$NS searches.

Recently, the CONUS+ Collaboration reported the most precise reactor-based CE$\nu$NS measurement to date~\cite{Ackermann:2026mms,Ackermann:2026bup}. Operated at the Leibstadt nuclear power plant in Switzerland as the successor to the CONUS experiment~\cite{CONUS:2020skt}, the CONUS+ Run~2 campaign achieved an approximately fourfold increase in statistics, accumulating $1296\;\mathrm{kg\cdot days}$ of reactor-on exposure across three high-purity germanium detectors labeled C3, C6, and C7. 

Analyzing the region of interest (ROI) up to $800~\mathrm{eV}_{ee}$ with detector-specific sub-keV energy thresholds, CONUS+ Run~2 measured the reactor-correlated CE$\nu$NS signal with a relative precision of $9.9\%$, consistent with the SM prediction. Building upon the earlier CONUS and CONUS+ Run~1 datasets—which already spurred extensive studies of nuclear physics~\cite{Rossi:2023brv,DeRomeri:2026ge}, the low-energy weak mixing angle~\cite{AtzoriCorona:2025ygn,Alpizar-Venegas:2025wor,DeRomeri:2026ge}, and various BSM scenarios~\cite{Alpizar-Venegas:2025wor,CONUS:2021dwh,DeRomeri:2025csu,Chattaraj:2025fvx,Benavides:2026conus}—the enhanced precision of Run~2 offers a timely opportunity to substantially tighten constraints on new neutrino interactions.

In this work, we perform a combined statistical analysis of the CONUS+ Run~2 datasets using a binned Poisson likelihood to derive updated constraints on three prominent BSM frameworks: light mediators, neutrino electromagnetic properties, and non-standard neutrino interactions (NSIs). For light mediators, we investigate both a universal light scalar ($\phi$) and four light vector ($Z'$) scenarios—namely, the universal vector model and the anomaly-free $U(1)_{B-L}$, $U(1)_{B-2L_e-L_\mu}$, and $U(1)_{B-3L_e}$ gauge extensions. For the electromagnetic sector, we constrain the effective electron neutrino magnetic moment, charge radius, and electric millicharge, while for NSIs we examine both one-parameter limits and two-dimensional correlations across flavor-conserving and flavor-changing couplings. Our results provide a direct update to previous CONUS+ Run~1 limits and highlight the complementarity of high-precision reactor CE$\nu$NS measurements with spallation-source, solar, direct detection, and accelerator probes.

The remainder of this paper is organized as follows. In Sec.~\ref{sec:theo}, we review the theoretical formalism for CE$\nu$NS within the SM and outline the BSM frameworks considered in this study. Section~\ref{sec:data_analysis} details the CONUS+ Run~2 experimental setup, event-rate modeling, and statistical analysis procedure. Our predicted event spectra, derived constraints, and comparisons with existing bounds are presented in Sec.~\ref{sec:results}. Finally, we summarize our main findings and conclude in Sec.~\ref{sec:conc}.

\section{Theoretical Framework}
\label{sec:theo}

\subsection{CE$\nu$NS in the Standard Model}
\label{subsec:sm_cevns}

Within the SM, the differential cross section for coherent elastic neutrino--nucleus scattering (CE$\nu$NS) on a target isotope $i$ of mass $M_i$, proton number $Z$, and neutron number $N_i$ ($A_i = Z + N_i$) with respect to the nuclear recoil kinetic energy $T_{\mathcal{N}}$ is given by \cite{Freedman:1973yd}
\begin{align}
    \left(\frac{d\sigma_i}{dT_{\mathcal{N}}}\right)_{\mathrm{SM}} =
    \frac{G_F^2 M_i}{\pi}\, Q_{\mathrm{SM},i}^2
    \Bigg(1 & - \frac{M_i T_{\mathcal{N}}}{2 E_\nu^2}
    - \frac{T_{\mathcal{N}}}{E_\nu}\Bigg) F_i^2(q^2),
    \label{eq:cevns_difcross}
\end{align}
where $G_F$ is the Fermi constant and $E_\nu$ is the incident neutrino energy. Because the interaction proceeds via neutral-current $Z$-boson exchange and depends quadratically on the weak charge, Eq.~\eqref{eq:cevns_difcross} holds identically for both neutrinos and antineutrinos. The coherent weak nuclear charge $Q_{\mathrm{SM},i}$ is defined as
\begin{equation}
    Q_{\mathrm{SM},i} = g_V^p Z + g_V^n N_i,
    \label{eq:weak_chrg}
\end{equation}
where the tree-level proton and neutron vector couplings are $g_V^p = \frac{1}{2} - 2\sin^2\theta_W$ and $g_V^n = -\frac{1}{2}$, respectively. Although the proton contribution is strongly suppressed owing to $\sin^2\theta_W \approx 1/4$, it introduces direct sensitivity to the weak mixing angle. 
Including low-energy electroweak radiative corrections in the $\overline{\mathrm{MS}}$ scheme~\cite{Erler:2013xha,AtzoriCorona:2024rtv}---together with the flavor-dependent neutrino charge radius contribution and the zero-momentum-transfer weak mixing angle $\sin^2\theta_W(0) = 0.23873(5)$ from the PDG~\cite{PDG}---the effective $\bar{\nu}_e$--nucleon couplings take the values $g_V^p(\nu_e) = 0.0379$ and $g_V^n = -0.5117$.
For natural germanium, only $^{73}\mathrm{Ge}$ ($7.75\%$ isotopic abundance) possesses a nonzero nuclear spin, so the spin-dependent axial-vector contribution is suppressed by $\mathcal{O}(1/A_i^2)$ relative to the coherent vector term and is safely neglected.

The nuclear form factor $F_i(q^2)$ accounts for the finite spatial distribution of nucleons and the resulting loss of coherence at momentum transfer $q = |\vec{q}| = \sqrt{2 M_i T_{\mathcal{N}}}$. Assuming identical proton and neutron distributions, $F_{p,i}(q^2) = F_{n,i}(q^2) \equiv F_i(q^2)$, we adopt the Klein--Nystrand parameterization~\cite{Klein:1999qj},
\begin{equation}
    F_i(q^2) = \frac{3\,j_1(q R_A)}{q R_A}\,\frac{1}{1 + (q a_k)^2},
    \label{eq:kn_ff}
\end{equation}
where $j_1(x)$ is the spherical Bessel function of the first kind of order one, $R_A = 1.23\,A_i^{1/3}\,\mathrm{fm}$ is the nuclear radius, and $a_k = 0.7\,\mathrm{fm}$ is the Yukawa range parameter. At reactor antineutrino energies ($E_\nu \lesssim 11~\mathrm{MeV}$), the typical momentum transfer in germanium remains modest ($q \sim 10\text{--}22~\mathrm{MeV}$, corresponding to $qR_A \lesssim 0.6$), ensuring $F_i^2(q^2) \gtrsim 0.90$ and minimal sensitivity to the specific choice of nuclear form-factor parameterization.

\subsection{Light Mediators}
\label{subsec:light_mediators}

When the mass of a new mediator is comparable to or smaller than the typical momentum transfer in reactor CE$\nu$NS ($m_{\mathrm{med}} \lesssim q \sim \mathcal{O}(10\,\mathrm{MeV})$), the four-fermion contact approximation breaks down and the mediator propagator must be retained explicitly~\cite{Cerdeno:2016sfi,Lindner:2016wff,CONUS:2021dwh,Coloma:2022avw}.
In this work, we analyze five benchmark simplified models: four light vector mediators---a flavor-universal vector $Z'$ and three anomaly-free gauged $U(1)'$ extensions, namely $U(1)_{B-L}$, $U(1)_{B-2L_e-L_\mu}$, and $U(1)_{B-3L_e}$ (for a comprehensive review of $U(1)'$ gauge bosons, see Ref.~\cite{Langacker:2008yv})---as well as a light scalar mediator $\phi$. Note that while these anomaly-free $U(1)'$ extensions also couple to electrons, the present analysis is restricted to the pure CE$\nu$NS channel, thereby yielding conservative nuclear-recoil bounds.

\textit{Light vector mediators.}---We consider a neutral vector boson $Z'$ of mass $m_{Z'}$ coupling to electron neutrinos and first-generation quarks via the interaction Lagrangian
\begin{equation}
    \mathcal{L}_{Z'} = Z'_\mu \left(
        g_{Z'}^{\nu_e}\,\bar{\nu}_e \gamma^\mu P_L \nu_e
        + \sum_{q=u,d} g_{Z'}^q\,\bar{q} \gamma^\mu q
    \right),
    \label{eq:zprime_lagrangian}
\end{equation}
where $g_{Z'}^{\nu_e}$ and $g_{Z'}^q$ denote the vector couplings to electron neutrinos and first-generation quarks ($q = u, d$), respectively. For the universal vector benchmark ($g_{Z'}^u = g_{Z'}^d \equiv g_{Z'}^q$), the CE$\nu$NS cross section depends on these couplings only through their product, allowing us to express our bounds in terms of the effective coupling $g_{Z'} \equiv \sqrt{g_{Z'}^{\nu_e} g_{Z'}^q}$. For the anomaly-free $U(1)'$ gauge extensions---namely $U(1)_{B-L}$, $U(1)_{B-2L_e-L_\mu}$, and $U(1)_{B-3L_e}$~\cite{Heeck:2018nzc,Coloma:2020gfv}---the fermion couplings are fixed by the single gauge coupling $g_{Z'}$ and the model-dependent charges as $g_{Z'}^f = g_{Z'} Q'_f$, as summarized in Table~\ref{tab:u1_charges}.

Because vector exchange preserves neutrino helicity, the $Z'$ amplitude interferes coherently with SM $Z$-boson exchange, modifying the differential CE$\nu$NS cross section in Eq.~\eqref{eq:cevns_difcross} via the recoil-energy-dependent replacement $Q_{\mathrm{SM},i} \to Q_{V,i}(T_{\mathcal{N}})$:
\begin{equation}
    Q_{V,i}(T_{\mathcal{N}}) = Q_{\mathrm{SM},i}
    + \frac{g_{Z'}^2\, Q'_{\nu_e}\, Q'_{\mathcal{N},i}}{\sqrt{2}\,G_F \left(2 M_i T_{\mathcal{N}} + m_{Z'}^2\right)},
    \label{eq:zprime_charge}
\end{equation}
where $Q'_{\mathcal{N},i} = (2 Q'_u + Q'_d) Z + (Q'_u + 2 Q'_d) N_i$ is the coherent $Z'$ charge of isotope $i$. Under charge conjugation, both the SM and $Z'$ antineutrino vertices flip sign simultaneously, preserving the relative sign in Eq.~\eqref{eq:zprime_charge} for reactor $\bar{\nu}_e$ scattering.

\begin{table}[t]
    \centering
    \caption{Charge assignments and effective nuclear charge products $Q'_{\nu_e} Q'_{\mathcal{N},i}$ for the light vector mediator models considered in this work ($A_i = Z + N_i$).}
    \label{tab:u1_charges}
    \begin{ruledtabular}
    \begin{tabular}{lccc}
        Model & $Q'_u = Q'_d$ & $Q'_{\nu_e}$ & $Q'_{\nu_e} Q'_{\mathcal{N},i}$ \\
        \hline
        $\mathrm{Universal}$ & $+1$ & $+1$ & $+3A_i$ \\
        $U(1)_{B-L}$ & $+1/3$ & $-1$ & $-A_i$ \\
        $U(1)_{B-2L_e-L_\mu}$ & $+1/3$ & $-2$ & $-2A_i$ \\
        $U(1)_{B-3L_e}$ & $+1/3$ & $-3$ & $-3A_i$ \\
    \end{tabular}
    \end{ruledtabular}
\end{table}

As seen in Table~\ref{tab:u1_charges}, the universal vector model produces a positive shift ($Q'_{\nu_e} Q'_{\mathcal{N},i} > 0$) that opposes $Q_{\mathrm{SM},i} < 0$, inducing destructive interference and a characteristic degeneracy corridor in the $(m_{Z'}, g_{Z'})$ plane where $Q_{V,i}(T_{\mathcal{N}}) \approx -Q_{\mathrm{SM},i}$ reproduces the SM rate. By contrast, all three anomaly-free $U(1)'$ models yield $Q'_{\nu_e} Q'_{\mathcal{N},i} < 0$, leading to constructive interference across the entire parameter space. Moreover, since short-baseline reactor CE$\nu$NS probes exclusively $\bar{\nu}_e$, the coupling bounds for $U(1)_{B-2L_e-L_\mu}$ (equivalent to $U(1)_{B-2L_e-L_\tau}$ in this channel) and $U(1)_{B-3L_e}$ scale relative to $U(1)_{B-L}$ by factors of $1/\sqrt{2}$ and $1/\sqrt{3}$, respectively. In all four models, Eq.~\eqref{eq:zprime_charge} smoothly interpolates between the mass-independent light-mediator plateau ($m_{Z'} \ll q$, enhanced as $1/T_{\mathcal{N}}$) and the contact-interaction regime ($m_{Z'} \gg q$, scaling as $g_{Z'}/m_{Z'} = \mathrm{const.}$).

\textit{Light scalar mediator.}---We also consider a real light scalar boson $\phi$ of mass $m_\phi$ interacting with neutrinos and first-generation quarks through
\begin{equation}
    \mathcal{L}_\phi = -\phi \left(
        \sum_{q=u,d} g_\phi^q \bar{q} q
        + g_\phi^{\nu_\ell} \bar{\nu}_{\ell R} \nu_{\ell L} +\mathrm{h.c.}
    \right),
    \label{eq:scalar_lagrangian}
\end{equation}
where $g_\phi^q$ and $g_\phi^{\nu_\ell}$ denote the scalar couplings to first-generation quarks ($q = u, d$) and neutrinos of flavor $\ell$, respectively. Assuming universal quark couplings ($g_\phi^u = g_\phi^d \equiv g_\phi^q$), the CE$\nu$NS cross section depends on the individual couplings only through their product, allowing us to express our bounds in terms of the effective coupling
$ g_\phi \equiv \sqrt{g_\phi^{\nu_\ell} g_\phi^q}$.
Since scalar exchange flips neutrino helicity, it does not interfere with the helicity-conserving SM amplitude and adds incoherently to Eq.~\eqref{eq:cevns_difcross}~\cite{Cerdeno:2016sfi}:
\begin{equation}
    \left(\frac{d\sigma_i}{dT_{\mathcal{N}}}\right)_\phi =
    \frac{M_i^2 T_{\mathcal{N}}}{4\pi E_\nu^2}\,
    \frac{g_\phi^4\, Q_{\phi,i}^2}{(2 M_i T_{\mathcal{N}} + m_\phi^2)^2}\,
    F_i^2(q^2).
    \label{eq:scalar_difcross}
\end{equation}
The coherent scalar nuclear charge $Q_{\phi,i}$ is expressed in terms of the nucleon scalar quark form factors $f_{T_q}^N$ as
\begin{equation}
    Q_{\phi,i} = Z \sum_{q=u,d} \frac{m_p}{m_q} f_{T_q}^p
    + N_i \sum_{q=u,d} \frac{m_n}{m_q} f_{T_q}^n,
    \label{eq:scalar_charge}
\end{equation}
where we adopt the hadronic form-factor values~\cite{DelNobile:2021wmp}
\begin{equation}
    f_{T_u}^p = 0.026, \quad f_{T_d}^p = 0.038, \\ \quad f_{T_u}^n = 0.018, \quad f_{T_d}^n = 0.056,
    \label{eq:hadronic_ff}
\end{equation}
together with the light-quark masses $m_u$ and $m_d$ from the PDG~\cite{PDG}. Owing to the purely additive $g_\phi^4$ dependence and the $T_{\mathcal{N}}/(2 M_i T_{\mathcal{N}} + m_\phi^2)^2$ kinematic factor in Eq.~\eqref{eq:scalar_difcross}, a light scalar mediator strictly enhances the low-energy recoil spectrum, particularly in the sub-MeV mass regime where the cross section scales as $1/T_{\mathcal{N}}$.

\subsection{Non-standard Neutrino Interactions}
\label{subsec:nsi}

In the heavy-mediator regime, where the mass of the new-physics mediator greatly exceeds the typical momentum transfer in reactor CE$\nu$NS ($m_{\mathrm{med}} \gg q \sim \mathcal{O}(10\,\mathrm{MeV})$), new neutral-current interactions between neutrinos and first-generation quarks can be described in a model-independent manner by dimension-six four-fermion effective operators~\cite{Wolfenstein:1977ue,Barranco:2005yy,Farzan:2017xzy}. The corresponding neutral-current nonstandard interaction (NSI) Lagrangian is given by
\begin{align}
    \mathcal{L}_{\mathrm{NSI}}^{\mathrm{NC}} = -2\sqrt{2}\,G_F
    \sum_{\alpha,\beta} \sum_{\substack{f=u,d  \\ P=P_L,P_R}} & \varepsilon_{\alpha\beta}^{fP}
    \left(\bar{\nu}_\alpha \gamma^\mu P_L \nu_\beta\right) \nonumber \\
    & \times \left(\bar{f} \gamma_\mu P f\right),
    \label{eq:nsi_lagrangian}
\end{align}
where $\alpha, \beta \in \{e, \mu, \tau\}$ denote neutrino flavors, $f \in \{u, d\}$ runs over valence up and down quarks, $P \in \{P_L, P_R\}$ with $P_{L,R} = (1 \mp \gamma_5)/2$ are the chirality projection operators, and $\varepsilon_{\alpha\beta}^{fP}$ are dimensionless parameters quantifying the strength of the new interactions relative to $G_F$. Hermiticity of the Lagrangian requires $\varepsilon_{\beta\alpha}^{fP} = (\varepsilon_{\alpha\beta}^{fP})^*$.

For coherent elastic scattering on a nucleus, the vector couplings $\varepsilon_{\alpha\beta}^{fV} \equiv \varepsilon_{\alpha\beta}^{fL} + \varepsilon_{\alpha\beta}^{fR}$ add coherently over all constituent nucleons, whereas the axial-vector combinations $\varepsilon_{\alpha\beta}^{fA} \equiv \varepsilon_{\alpha\beta}^{fR} - \varepsilon_{\alpha\beta}^{fL}$ couple to the nuclear spin and are strongly suppressed in natural germanium. Because the four-fermion vector NSI operators share the exact same Lorentz and recoil-energy structure as SM $Z$-boson exchange, their inclusion modifies the differential CE$\nu$NS cross section in Eq.~\eqref{eq:cevns_difcross} solely through the replacement $Q_{\mathrm{SM},i}^2 \to Q_{\mathrm{NSI},i}^2$. For reactor electron antineutrinos ($\bar{\nu}_e$) scattering off a germanium isotope $i$ with $Z$ protons and $N_i$ neutrons, the flavor-conserving couplings ($\varepsilon_{ee}^{uV}, \varepsilon_{ee}^{dV}$) interfere coherently with the SM amplitude, while the flavor-changing transitions ($\bar{\nu}_e \to \bar{\nu}_\mu, \bar{\nu}_\tau$) lead to distinguishable final states and contribute incoherently. Accounting for the quark content of protons ($uud$) and neutrons ($udd$), the effective squared weak charge reads
\begin{align}
    Q_{\mathrm{NSI},i}^2 =\; &\Big[ Q_{\mathrm{SM},i}
    + \varepsilon_{ee}^{uV}(2Z + N_i)
    + \varepsilon_{ee}^{dV}(Z + 2N_i) \Big]^2 \nonumber \\
    &+ \sum_{\alpha=\mu,\tau} \Big| \varepsilon_{e\alpha}^{uV}(2Z + N_i)
    + \varepsilon_{e\alpha}^{dV}(Z + 2N_i) \Big|^2.
    \label{eq:nsi_charge}
\end{align}

In this work, we treat all NSI couplings as real parameters and evaluate Eq.~\eqref{eq:nsi_charge} explicitly for each of the five natural germanium isotopes before summing over isotopic abundances in Eq.~\eqref{eq:diff_rate_nr}. 
Because the relative sign between $Q_{\mathrm{SM},i} < 0$ and the flavor-conserving NSI terms determines the nature of the interference, negative (positive) values of $\epsilon_{ee}^{fV}$ enhance (suppress) the CE$\nu$NS rate relative to the SM. Destructive interference gives rise to an approximately linear degeneracy band in the $(\epsilon_{ee}^{uV}, \epsilon_{ee}^{dV})$ plane along $Q_{\mathrm{NSI},i} \approx -Q_{\mathrm{SM},i}$, where the SM event rate is reproduced, whereas the zero-rate line $Q_{\mathrm{NSI},i} \approx 0$ lying between the two allowed bands is excluded by the data. Since natural germanium comprises five stable isotopes with slightly different neutron-to-proton ratios $N_i/Z$, the exact cancellation condition cannot be satisfied simultaneously across all isotopes, rendering the degeneracy only approximate. 
By contrast, flavor-changing couplings $\varepsilon_{e\mu}^{fV}$ and $\varepsilon_{e\tau}^{fV}$ enter strictly quadratically in Eq.~\eqref{eq:nsi_charge}, producing identical, purely constructive enhancements of the recoil rate and yielding symmetric constraints around zero.

\subsection{Neutrino Electromagnetic Properties}
\label{subsec:em_props}

Although neutrinos are electrically neutral at tree level in the SM, quantum loop corrections and physics beyond the SM can induce effective electromagnetic interactions~\cite{Giunti:2014ixa}. In low-energy CE$\nu$NS experiments, these interactions are probed through three main electromagnetic parameters: the neutrino magnetic moment, the neutrino charge radius, and a possible neutrino electric millicharge. In our analysis of the CONUS+ Run~2 reactor $\bar{\nu}_e$ data, each electromagnetic parameter is constrained individually within the coherent nuclear-recoil channel.

\textit{Neutrino magnetic moment.}---A nonzero neutrino magnetic (or electric dipole) moment flips the neutrino helicity during scattering. Because this helicity-flipping amplitude does not interfere with the helicity-conserving SM weak interaction, its contribution adds incoherently to the SM differential cross section in Eq.~\eqref{eq:cevns_difcross}~\cite{Vogel:1989iv}:
\begin{align}
    \left(\frac{d\sigma_i}{dT_{\mathcal{N}}}\right)_{\mathrm{MM}} =
    \frac{\pi \alpha_{\mathrm{em}}^2}{m_e^2}
    \left(\frac{1}{T_{\mathcal{N}}} - \frac{1}{E_\nu}\right) & Z^2
    \left|\frac{\mu_{\nu_e}^{\mathrm{eff}}}{\mu_{\mathrm{B}}}\right|^2 \nonumber \\
    & \times F_i^2(q^2),
    \label{eq:mm_difcross}
\end{align}
where $\alpha_{\mathrm{em}}$ is the fine-structure constant, $m_e$ is the electron mass, $\mu_{\mathrm{B}} = e/(2 m_e)$ is the Bohr magneton, and $\mu_{\nu_e}^{\mathrm{eff}}$ denotes the effective magnetic moment of reactor electron antineutrinos at short baseline (encapsulating both diagonal and transition magnetic and electric dipole moments). Owing to the $1/T_{\mathcal{N}}$ enhancement in Eq.~\eqref{eq:mm_difcross}, sub-keV-threshold germanium detectors are particularly sensitive to $\mu_{\nu_e}^{\mathrm{eff}}$ near the low-energy threshold. In the minimal extension of the SM with massive Dirac neutrinos, the predicted magnetic moment is exceedingly small, $\mu_\nu^{\mathrm{SM}} \approx 3.2 \times 10^{-19}\,(m_\nu/\mathrm{eV})\,\mu_{\mathrm{B}}$~\cite{Giunti:2014ixa}; hence, any measurable signal at current experimental sensitivities would provide unambiguous evidence for BSM physics.

\textit{Neutrino charge radius.}---Even for strictly neutral neutrinos, electroweak radiative corrections generate a finite, gauge-invariant charge radius~\cite{Bernabeu:2000hf,Bernabeu:2002nw}. For electron neutrinos, the SM prediction is $\langle r_{\nu_e}^2 \rangle_{\mathrm{SM}} = -0.83 \times 10^{-32}\,\mathrm{cm}^2$~\cite{Bernabeu:2002nw,Giunti:2014ixa}. Unlike the magnetic moment, the charge radius preserves neutrino helicity and therefore interferes coherently with the SM weak amplitude. Focusing on the flavor-diagonal charge radius $\langle r_{\nu_e}^2 \rangle$, its effect is incorporated into Eq.~\eqref{eq:cevns_difcross} through a shift in the proton weak coupling in Eq.~\eqref{eq:weak_chrg}~\cite{AtzoriCorona:2024rtv}:
\begin{equation}
    g_V^p(\nu_e) \to g_V^p(\nu_e)
    - \frac{\sqrt{2}\pi\alpha_{\mathrm{em}}}{3 G_F}
    \Big(\langle r_{\nu_e}^2 \rangle - \langle r_{\nu_e}^2 \rangle_{\mathrm{SM}}\Big),
    \label{eq:cr_shift}
\end{equation}
where the prefactor satisfies $\sqrt{2}\pi\alpha_{\mathrm{em}}/(3G_F) \simeq \frac{2}{3} m_W^2 \sin^2\theta_W$ at tree level. 
Because the radiative-corrected value $g_V^p(\nu_e) = 0.0379$ adopted in Sec.~\ref{subsec:sm_cevns} already includes the SM charge-radius contribution, subtracting $\langle r_{\nu_e}^2 \rangle_{\mathrm{SM}}$ in Eq.~\eqref{eq:cr_shift} avoids double counting and ensures that the SM cross section is exactly recovered at $\langle r_{\nu_e}^2 \rangle = \langle r_{\nu_e}^2 \rangle_{\mathrm{SM}}$. Furthermore, since both the weak neutral-current coupling and the charge radius reverse sign under charge conjugation ($\langle r_{\bar{\nu}_e}^2 \rangle = -\langle r_{\nu_e}^2 \rangle$), the interference structure in Eq.~\eqref{eq:cr_shift} remains identical for reactor $\bar{\nu}_e$ when expressed in terms of $\langle r_{\nu_e}^2 \rangle$. Since $Q_{\mathrm{SM},i} < 0$, shifts with $\langle r_{\nu_e}^2 \rangle < \langle r_{\nu_e}^2 \rangle_{\mathrm{SM}}$ ($\langle r_{\nu_e}^2 \rangle > \langle r_{\nu_e}^2 \rangle_{\mathrm{SM}}$) induce destructive (constructive) interference with the SM near the baseline value, producing an asymmetric $\chi^2$ profile around the SM point.

\textit{Neutrino electric millicharge.}---In several BSM extensions with hidden gauge symmetries or broken electric charge quantization, neutrinos may acquire a tiny electric millicharge $q_{\nu_e}$~\cite{Giunti:2014ixa}. Single-photon exchange mediated by a diagonal millicharge is also helicity conserving and enters the CE$\nu$NS cross section in Eq.~\eqref{eq:cevns_difcross} via a momentum-transfer-dependent shift of the proton coupling in Eq.~\eqref{eq:weak_chrg}:
\begin{equation}
    g_V^p(\nu_e) \to g_V^p(\nu_e)
    + \frac{\sqrt{2}\pi\alpha_{\mathrm{em}}}{G_F M_i T_{\mathcal{N}}}\, q_{\nu_e},
    \label{eq:mc_shift}
\end{equation}
As in the charge-radius case, CPT invariance ($q_{\bar{\nu}_e} = -q_{\nu_e}$) preserves the relative sign between the SM and millicharge amplitudes for reactor $\bar{\nu}_e$ scattering when written in terms of $q_{\nu_e}$. Because of the $1/q^2 \propto 1/T_{\mathcal{N}}$ photon-propagator scaling in Eq.~\eqref{eq:mc_shift}, the millicharge induces a pronounced spectral distortion at the lowest recoil energies, interfering destructively with the SM for $q_{\nu_e} > 0$ and constructively for $q_{\nu_e} < 0$.

\section{Data Analysis Details}
\label{sec:data_analysis}
\subsection{CONUS+ Run~2 Dataset and Signal Modeling}
\label{subsec:dataset_response}
Our analysis is based on the Run~2 dataset of the CONUS+ experiment~\cite{Ackermann:2026mms,Ackermann:2026bup}, which operates high-purity point-contact germanium (HPGe) detectors at short baseline from the core of a commercial boiling water reactor. Specifically, we analyze the data collected by the three low-threshold detectors $\mathcal{D}\equiv\{\mathrm{C3},\mathrm{C6},\mathrm{C7}\}$ across the ionization-energy region of interest (ROI) extending up to $E_{\mathrm{ee}}^{\max}=0.80~\mathrm{keV}_{\mathrm{ee}}$. The measured recoil spectra are binned uniformly with a bin width of $\Delta E_{\mathrm{ee}}=10~\mathrm{eV}_{\mathrm{ee}}$ above each detector's fit threshold: $E_{\mathrm{th}}=0.17~\mathrm{keV}_{\mathrm{ee}}$ for C3 (63~bins), $0.16~\mathrm{keV}_{\mathrm{ee}}$ for C6 (64~bins), and $0.15~\mathrm{keV}_{\mathrm{ee}}$ for C7 (65~bins), yielding a total of $N_{\mathrm{bins}}=192$ energy bins and a combined reactor-on exposure of $1296~\mathrm{kg\cdot d}$. Although the collaboration adopts a nominal threshold of $140~\mathrm{eV}_{\mathrm{ee}}$ for C7~\cite{Ackermann:2026mms,Ackermann:2026bup}, the combined-fit spectra in Fig.~1 of Ref.~\cite{Ackermann:2026mms}---from which we extract the observed event counts $N^{(d)}_{\mathrm{obs},k}$ and the combined-fit total background predictions $B^{(d)}_{\mathrm{tot},k}$ (incorporating all simulated cosmogenic, radiogenic, and environmental background components~\cite{Ackermann:2026bup})---are shown only for $E_{\mathrm{ee}}\in[0.15,0.80]~\mathrm{keV}_{\mathrm{ee}}$. The $140$--$150~\mathrm{eV}_{\mathrm{ee}}$ bin of C7 is available only in the logarithmic individual-detector panel (Fig.~20 of Ref.~\cite{Ackermann:2026bup}); to keep all bins on a strictly uniform combined-fit footing, we start the C7 fit at $E_{\mathrm{th}}=0.15~\mathrm{keV}_{\mathrm{ee}}$. This choice also reduces sensitivity to the steep trigger-efficiency turn-on and residual energy-scale nonlinearity near threshold, yielding slightly more conservative BSM limits. Moreover, the collaboration's consistency check with all thresholds raised by $10~\mathrm{eV}_{\mathrm{ee}}$ leaves $N_s/N_{\mathrm{SM}}=0.92$ unchanged (Table~VI of Ref.~\cite{Ackermann:2026bup}), indicating that the measured ratio is insensitive to a small upward shift in the threshold. The key experimental characteristics of the three detectors are summarized in Table~\ref{tab:conus_detectors}.
\begin{table}[ht]
    \centering
    \caption{Summary of the CONUS+ Run~2 HPGe detectors included in the joint statistical analysis~\cite{Ackermann:2026bup}, listing the analysis threshold $E_{\mathrm{th}}$, upper ROI bound $E_{\mathrm{ee}}^{\max}$, number of $10\,\mathrm{eV_{ee}}$ bins $N_d$, active exposure $\mathcal{E}_d$, and electronic noise resolution $\mathrm{FWHM}_0$.}
    \label{tab:conus_detectors}
    \begin{ruledtabular}
    \begin{tabular}{lccccc}
        Det. & $E_{\mathrm{th}}$ & $E_{\mathrm{ee}}^{\max}$ & $N_d$ & $\mathcal{E}_d$ & $\mathrm{FWHM}_0$ \\
             & [$\mathrm{eV_{ee}}$] & [$\mathrm{eV_{ee}}$] & & [$\mathrm{kg\cdot d}$] & [$\mathrm{eV_{ee}}$] \\
        \hline
        C3 & 170 & 800 & 63 & 213.00 & 53.5 \\
        C6 & 160 & 800 & 64 & 544.00 & 52.1 \\
        C7 & 150 & 800 & 65 & 539.00 & 51.7 \\
        \hline
        Total & -- & -- & 192 & 1296.00 & -- \\
    \end{tabular}
    \end{ruledtabular}
\end{table}

To compute the expected CE$\nu$NS differential recoil rate per unit target mass, we sum over the five stable isotopes comprising natural germanium---$^{70}\mathrm{Ge}$ ($20.57\%$), $^{72}\mathrm{Ge}$ ($27.45\%$), $^{73}\mathrm{Ge}$ ($7.75\%$), $^{74}\mathrm{Ge}$ ($36.50\%$), and $^{76}\mathrm{Ge}$ ($7.73\%$)---weighted by their respective isotopic abundances $f_i$:
\begin{equation}
    \frac{dR}{dT_{\mathcal{N}}} = N_T \sum_{i} f_i \int_{E_\nu^{\min}}^{E_\nu^{\max}} dE_\nu \, \frac{d\Phi_{\bar{\nu}_e}}{dE_\nu} \, \frac{d\sigma_i}{dT_{\mathcal{N}}},
    \label{eq:diff_rate_nr}
\end{equation}
where $N_T$ is the number of germanium nuclei per kilogram of target mass, $T_{\mathcal{N}}$ is the nuclear recoil kinetic energy, and $E_\nu^{\min}(T_{\mathcal{N}}) = \frac{1}{2}\big(T_{\mathcal{N}} + \sqrt{T_{\mathcal{N}}^2 + 2 M_i T_{\mathcal{N}}}\big)$ is the minimum antineutrino energy required to induce a recoil of energy $T_{\mathcal{N}}$ on an isotope of mass $M_i$. 
For the reactor antineutrino flux $d\Phi_{\bar{\nu}_e}/dE_\nu$ produced by the $3.6\,\mathrm{GW}_{\mathrm{th}}$ (nominal) boiling water reactor at the Leibstadt nuclear power plant (KKL, $20.7\,\mathrm{m}$ baseline, total flux $\sim 10^{13}\,\mathrm{cm}^{-2}\,\mathrm{s}^{-1}$)~\cite{Ackermann:2026mms,Ackermann:2026bup}, we adopt the data-driven CONUS+ prescription~\cite{Ackermann:2026bup}. It is based on the unfolded Daya Bay reactor spectrum~\cite{DayaBay:2021dqj}, adjusted to the fission fractions of the KKL reactor during Run~2 ($^{235}$U, $^{238}$U, $^{239}$Pu, $^{241}$Pu: $0.54$, $0.08$, $0.31$, $0.07$) and extended up to $E_\nu^{\max} = 11\,\mathrm{MeV}$ with the Daya Bay high-energy measurement~\cite{DayaBay:2022prl}. The flux is computed for the average Run~2 thermal power of $3552\,\mathrm{MW}$, for which a conservative relative uncertainty of $1.5\%$ is assumed~\cite{Ackermann:2026bup}.
For the CONUS+ Run~2 analysis thresholds ($E_{\mathrm{th}} \geq 0.15~\mathrm{keV}_{ee}$, corresponding to $T_{\mathcal{N}} \gtrsim 0.9~\mathrm{keV}$), the kinematic threshold $E_\nu^{\min} \gtrsim 5.5~\mathrm{MeV}$ lies well above the $1.8~\mathrm{MeV}$ inverse beta decay threshold; consequently, the low-energy neutron-capture component does not contribute to the detectable recoil spectrum, and the integral in Eq.~\eqref{eq:diff_rate_nr}  is evaluated over $E_\nu \in [E_\nu^{\min}, 11~\mathrm{MeV}]$ using adaptive numerical quadrature with explicit knots at the tabulated Daya Bay energy nodes.

To fold the theoretical nuclear-recoil spectrum in Eq.~\eqref{eq:diff_rate_nr} with the detector response of each germanium crystal $d \in \mathcal{D}$, we first map the nuclear recoil energy $T_{\mathcal{N}}$ to the true electron-equivalent energy $E_{\mathrm{ee}}'(T_{\mathcal{N}})$ via the Lindhard quenching model~\cite{Lindhard:1963},
\begin{equation}
    E_{\mathrm{ee}}'(T_{\mathcal{N}}) = f_Q(T_{\mathcal{N}})\,T_{\mathcal{N}}, \qquad
    f_Q(T_{\mathcal{N}}) = \frac{k_{\mathrm{L}}\,g(\epsilon)}{1 + k_{\mathrm{L}}\,g(\epsilon)},
    \label{eq:lindhard}
\end{equation}
where $g(\epsilon) = 3\epsilon^{0.15} + 0.7\epsilon^{0.6} + \epsilon$ and $\epsilon = 11.5\,Z^{-7/3}\,(T_{\mathcal{N}}/\mathrm{keV})$ for $Z = 32$. Notably, the CONUS+ observations agree well with the Lindhard prediction for the directly measured germanium quenching parameter $k_{\mathrm{L}} = 0.162 \pm 0.004$~\cite{Bonhomme:2022,Ackermann:2026mms,Ackermann:2026bup}, for which we adopt the central value $k_{\mathrm{L}} = 0.162$.

In addition, we incorporate the detector-specific residual energy-scale nonlinearity $\Delta E_{\mathrm{NL}}^{(d)}(E_{\mathrm{ee}}')$ measured via pulser calibrations in CONUS+ Run~2~\cite{Ackermann:2026mms,Ackermann:2026bup} by linearly interpolating the tabulated calibration curves for each detector, where the shift reaches approximately $-3.5$ to $-4.7\,\mathrm{eV_{ee}}$ in the sub-$200\,\mathrm{eV_{ee}}$ region and tapers to $\lesssim 1\,\mathrm{eV_{ee}}$ above $250\,\mathrm{eV_{ee}}$. The expected CE$\nu$NS signal count $S_k^{(d)}(\vec{\theta})$ in the $k$th $10\,\mathrm{eV_{ee}}$ bin $[E_{k,\min}^{(d)}, E_{k,\max}^{(d)}]$ (from the detector threshold $E_{\mathrm{th},d}$ up to $800\,\mathrm{eV_{ee}}$) for a given BSM parameter set $\vec{\theta}$ is then computed via adaptive numerical integration directly over $T_{\mathcal{N}}$ and the reconstructed energy $E_{\mathrm{ee}}$:
\begin{align}
    S_k^{(d)}(\vec{\theta}) =\; &\mathcal{E}_d \int_{E_{k,\min}^{(d)}}^{E_{k,\max}^{(d)}} dE_{\mathrm{ee}}\,\varepsilon_{\mathrm{trig}}^{(d)}(E_{\mathrm{ee}})
    \int dT_{\mathcal{N}}\,\frac{dR(\vec{\theta})}{dT_{\mathcal{N}}} \nonumber \\
    &\times \frac{1}{\sqrt{2\pi}\,\sigma_d(E_{\mathrm{ee}}')}
    \exp\!\left\{-\frac{\big[E_{\mathrm{ee}} - \bar{E}_{\mathrm{ee}}^{(d)}(T_{\mathcal{N}})\big]^2}{2\sigma_d^2(E_{\mathrm{ee}}')}\right\},
    \label{eq:binned_signal}
\end{align}
where $\mathcal{E}_d$ is the reactor-on exposure, $\bar{E}_{\mathrm{ee}}^{(d)}(T_{\mathcal{N}}) = E_{\mathrm{ee}}'(T_{\mathcal{N}}) + \Delta E_{\mathrm{NL}}^{(d)}(E_{\mathrm{ee}}')$ is the nonlinearity-shifted mean reconstructed energy, and $\varepsilon_{\mathrm{trig}}^{(d)}(E_{\mathrm{ee}}) = \frac{1}{2}\big[1 + \mathrm{erf}\big((E_{\mathrm{ee}} - t_{1,d})/t_{2,d}\big)\big]$ is the pulser-calibrated trigger efficiency taken from Ref.~\cite{Ackermann:2026bup} (with $t_{1,d}$ and $t_{2,d}$ denoting the $50\%$ efficiency point and transition width, respectively). The energy-dependent Gaussian resolution width $\sigma_d(E_{\mathrm{ee}}')$ is parameterized as~\cite{Ackermann:2026bup}
\begin{equation}
    \sigma_d(E_{\mathrm{ee}}') = \sqrt{\sigma_{0,d}^2 + F_d\,\epsilon_{\mathrm{eh}}\,E_{\mathrm{ee}}' + \big(\beta_{\mathrm{res},d}\,E_{\mathrm{ee}}'\big)^2},
    \label{eq:resolution}
\end{equation}
combining the electronic noise term $\sigma_{0,d} = \mathrm{FWHM}_{0,d}/(2\sqrt{2\ln 2})$, the Fano contribution (with Fano factor $F_d$ and electron--hole pair creation energy $\epsilon_{\mathrm{eh}} = 2.96\,\mathrm{eV}$ in germanium), and the linear energy-scaling term $\beta_{\mathrm{res},d}$. 
All uncertainties associated with the detector response and signal prediction---including the Lindhard quenching parameter and the $\pm 3~\mathrm{eV}_{\mathrm{ee}}$ energy-scale uncertainty at threshold~\cite{Ackermann:2026bup}---are subsumed into the correlated signal normalization systematic $\sigma_\alpha$ introduced in Sec.~III\,B.

\subsection{Statistical Method}
\label{subsec:stat_framework}

To derive constraints on light mediators, nonstandard interactions, and neutrino electromagnetic properties on a strictly uniform statistical footing, we perform a simultaneous binned joint fit to all three CONUS+ Run~2 detectors ($d \in \mathcal{D}$, 192 bins in total) using the Poisson likelihood-ratio test statistic~\cite{Baker:1983tu}:
\begin{widetext}
	\begin{equation}
		\chi^2(\vec{\theta}) = \min_{\alpha,\,\vec{\beta}} \Bigg\{
		2 \sum_{d \in \mathcal{D}} \sum_{k=1}^{N_d} \bigg[
		N_{\mathrm{exp},k}^{(d)} - N_{\mathrm{obs},k}^{(d)}
		+ N_{\mathrm{obs},k}^{(d)} \ln\!\left(\frac{N_{\mathrm{obs},k}^{(d)}}{N_{\mathrm{exp},k}^{(d)}}\right)
		\bigg]
		+ \left(\frac{\alpha}{\sigma_\alpha}\right)^2
		+ \sum_{d \in \mathcal{D}} \left(\frac{\beta_d}{\sigma_\beta}\right)^2
		\Bigg\},
		\label{eq:baker_cousins_chi2}
	\end{equation}
\end{widetext}
where the expected total event count in bin $k$ of detector $d$ is modeled as
\begin{equation}
	N_{\mathrm{exp},k}^{(d)}(\vec{\theta}; \alpha, \beta_d) = (1 + \alpha)\,S_k^{(d)}(\vec{\theta}) + (1 + \beta_d)\,B_{\mathrm{tot},k}^{(d)},
	\label{eq:nexp_bin}
\end{equation}
and the logarithmic term in Eq.~\eqref{eq:baker_cousins_chi2} vanishes for any bin with $N_{\mathrm{obs},k}^{(d)} = 0$.
Here, $\alpha$ is a correlated signal normalization nuisance parameter with Gaussian prior width $\sigma_\alpha = 11.6\%$, matching the total relative uncertainty on the CONUS+ Run~2 SM signal prediction~\cite{Ackermann:2026bup}. This budget combines in quadrature the threshold energy-scale uncertainty ($8.0\%$, corresponding to the adopted $\pm 3~\mathrm{eV}_{\mathrm{ee}}$ systematic across all three detectors), the Lindhard quenching-factor uncertainty ($7.5\%$), the CE$\nu$NS cross-section uncertainty ($3.1\%$), and the reactor antineutrino flux and spectrum uncertainty ($2.1\%$), together with sub-percent contributions from the active mass ($0.7\%$) and trigger efficiency ($0.4\%$) (see Fig.~17 of Ref.~\cite{Ackermann:2026bup} and Fig.~2 of Ref.~\cite{Ackermann:2026mms}).
The three uncorrelated nuisance parameters $\vec{\beta}=(\beta_{\mathrm{C3}},\beta_{\mathrm{C6}},\beta_{\mathrm{C7}})$ allow for detector-specific background normalization shifts with Gaussian prior width $\sigma_\beta = 2.0\%$, chosen such that---together with the background-dominated control region---the induced uncertainty on the extracted signal yield ($\simeq 4\%$) is comparable to the $3.5\%$ background-model systematic quoted by the collaboration~\cite{Ackermann:2026bup}. 
At every point $\vec{\theta}$ in the BSM parameter space, the four nuisance parameters $(\alpha,\beta_{\mathrm{C3}},\beta_{\mathrm{C6}},\beta_{\mathrm{C7}})$ are profiled numerically while preserving physical positivity ($N_{\mathrm{exp},k}^{(d)} > 0$).

All parameter bounds in this work are extracted exclusively from the CE$\nu$NS channel. While elastic neutrino--electron scattering (E$\nu$ES) can in principle enhance sensitivity to neutrino electromagnetic properties and light mediators with electron couplings across the full $[0.15,\,0.80]~\mathrm{keV}_{\mathrm{ee}}$ window, tabulated component-by-component background templates, nuisance priors, and the $0.8$--$28~\mathrm{keV}_{\mathrm{ee}}$ sideband spectra used to constrain the individual background normalizations are not publicly available~\cite{Ackermann:2026mms,Ackermann:2026bup}. Modeling only the summed background $B_{\mathrm{tot},k}^{(d)}$ leaves broad-band E$\nu$ES searches susceptible to the residual background-shape uncertainties above the CE$\nu$NS signal region ($E_{\mathrm{ee}} \gtrsim 0.35~\mathrm{keV}_{\mathrm{ee}}$) as well as to low-energy atomic-ionization modeling uncertainties in germanium. Restricting the analysis uniformly to the CE$\nu$NS channel therefore ensures conservative, robust, and mutually consistent constraints across all three BSM sectors.

\begin{figure*}[ht]
	\centering
	\includegraphics[width=0.80\textwidth]{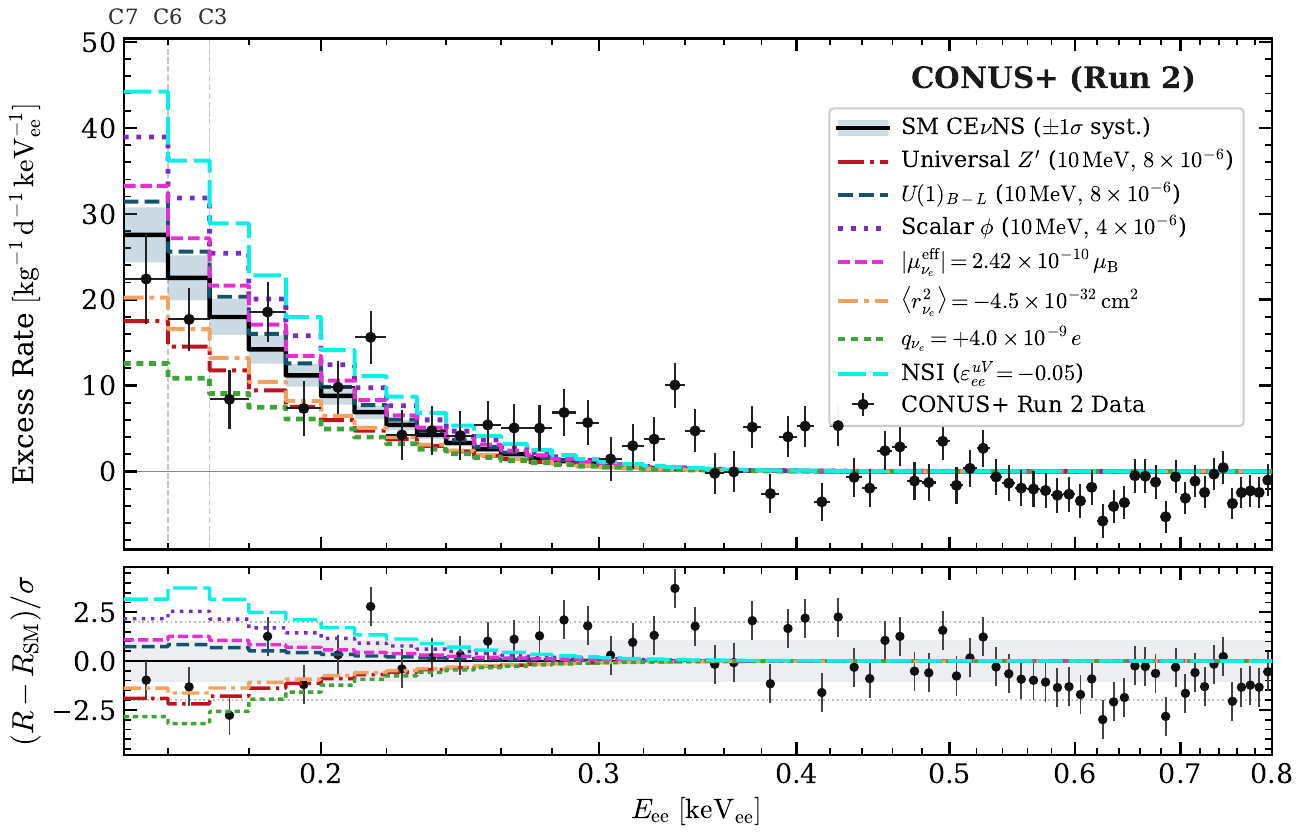}
	\caption{Combined exposure-weighted residual (background-subtracted) CE$\nu$NS spectrum of the CONUS+ Run~2 dataset (C3, C6, and C7) versus reconstructed electron-equivalent energy $E_{\mathrm{ee}}$ (top panel), along with the normalized residuals $(R - R_{\mathrm{SM}})/\sigma$ relative to the SM prediction (bottom panel). The solid black histogram and shaded band represent the SM expectation and its $\pm 1\sigma$ systematic uncertainty ($\sigma_\alpha = 11.6\%$), while colored histograms show representative BSM benchmark scenarios detailed in the text. 
	In the bottom panel, the gray band and horizontal dotted lines denote the $\pm 1\sigma$ and $\pm 2\sigma$ levels, respectively. The combined spectrum is shown for illustration; all statistical fits are performed simultaneously on the unmerged individual detector spectra.}
	\label{fig:conus_run2_residual_all_bsm}
\end{figure*}

Under the SM-only hypothesis, the simultaneous fit to the 192 bins yields $\chi^2_{\mathrm{SM}} = 292.6$ ($\chi^2_{\mathrm{SM}}/\mathrm{d.o.f.} = 1.52$), reflecting the mild background-shape mismatch above the signal region also noted in Ref.~\cite{Ackermann:2026bup}. The best-fit nuisance parameters ($\hat{\alpha} = -6.2\%$, $|\hat{\beta}_d| < 1\%$) lie well within their priors, corresponding to $1 + \hat{\alpha} \simeq 0.94$, consistent with $N_s/N_{\mathrm{SM}} = 0.92 \pm 0.14$~\cite{Ackermann:2026bup}. In Sec.~IV, all BSM constraints are derived from $\Delta\chi^2(\vec{\theta}) = \chi^2(\vec{\theta}) - \chi^2_{\min}$, where $\chi^2_{\min}$ is the global minimum for each scenario.

\section{Results}
\label{sec:results}
In this section, we present the constraints on physics beyond the SM derived from our simultaneous analysis of the CONUS+ Run~2 dataset (detectors C3, C6, and C7), probing reactor antineutrino ($\bar{\nu}_e$) CE$\nu$NS on natural germanium. Specifically, we investigate three BSM sectors: light vector and scalar mediators, neutrino electromagnetic properties, and vector nonstandard neutrino interactions (NSIs). Employing the $\chi^2$ formalism introduced in Sec.~\ref{sec:data_analysis}, we determine the allowed parameter spaces and exclusion bounds at the $90\%$ C.L. according to the relevant degrees of freedom, profiling over the correlated signal normalization and detector-specific background nuisance parameters at every point in parameter space. We begin by examining the measured recoil spectrum before presenting the resulting constraints for each BSM sector in turn.

To visually compare the observed CE$\nu$NS excess against the SM and representative BSM expectations across the full region of interest ($[0.15,\,0.80]\,\mathrm{keV_{ee}}$), Fig.~\ref{fig:conus_run2_residual_all_bsm} displays the combined exposure-weighted differential residual rate of the three CONUS+ Run~2 germanium detectors (C3, C6, and C7). Because the detectors operate with different low-energy thresholds---$E_{\mathrm{th}} = 0.15\,\mathrm{keV_{ee}}$ for C7, $0.16\,\mathrm{keV_{ee}}$ for C6, and $0.17\,\mathrm{keV_{ee}}$ for C3---the combined rate in each $10\,\mathrm{eV_{ee}}$ bin $k$ is constructed by summing only over the subset of active detectors $\mathcal{D}_k$:
\begin{equation}
    R_k = \frac{\sum_{d \in \mathcal{D}_k} \left[ N_{\mathrm{obs},k}^{(d)} - (1 + \hat{\beta}_d)\, B_{\mathrm{tot},k}^{(d)} \right]}{\Delta E_k \sum_{d \in \mathcal{D}_k} \mathcal{E}_d},
    \label{eq:combined_residual_rate}
\end{equation}
where $\mathcal{E}_d$ is the exposure of detector $d$, $\Delta E_k = 0.01\,\mathrm{keV_{ee}}$ is the bin width, and $\hat{\beta}_d$ are the best-fit profiled background shifts obtained from the global SM fit. The bin uncertainty combines the statistical and background contributions in quadrature, $\sigma_k = \sqrt{\sum_{d \in \mathcal{D}_k} \left[N_{\mathrm{obs},k}^{(d)} + (\sigma_\beta \hat{B}_k^{(d)})^2\right]} / (\Delta E_k \sum_{d \in \mathcal{D}_k} \mathcal{E}_d)$, with $\hat{B}_k^{(d)} = (1 + \hat{\beta}_d) B_{\mathrm{tot},k}^{(d)}$ and $\sigma_\beta = 2\%$. The theoretical SM and BSM histograms are exposure-averaged in the same manner. We stress that Eq.~\eqref{eq:combined_residual_rate} is used solely for visual presentation, whereas all parameter constraints in this work are derived from a simultaneous joint fit to the unmerged spectra of C3, C6, and C7 (192 bins in total).

As seen in Fig.~1, the measured excess agrees well with the SM prediction within uncertainties below $E_{\mathrm{ee}}\simeq 0.35~\mathrm{keV}_{\mathrm{ee}}$, where the CE$\nu$NS signal is concentrated. At higher energies, which serve as a background-dominated control region where all CE$\nu$NS contributions vanish, the residuals exhibit a mild wave-like structure at the $\lesssim 3\sigma$ level, indicative of the residual background-shape mismatch also noted by the collaboration (primarily in C3 and C6, without biasing the extracted signal)~\cite{Ackermann:2026bup} and reflected in $\chi^2_{\mathrm{SM}}/\mathrm{d.o.f.}=1.52$. To illustrate how different new-physics scenarios distort the recoil spectrum, we overlay representative benchmark curves spanning all three BSM sectors analyzed in this study:
(i) light mediators, namely the $U(1)_{B-L}$ vector ($m_{Z'} = 10\,\mathrm{MeV}$, $g = 8\times 10^{-6}$), universal vector $Z'$ ($m_{Z'} = 10\,\mathrm{MeV}$, $g = 8\times 10^{-6}$), and light scalar $\phi$ ($m_\phi = 10\,\mathrm{MeV}$, $g = 4\times 10^{-6}$);
(ii) neutrino electromagnetic properties, including an effective magnetic moment $|\mu_{\nu_e}^{\mathrm{eff}}| = 2.42\times 10^{-10}\,\mu_{\mathrm{B}}$, charge radius $\langle r_{\nu_e}^2 \rangle = -4.5\times 10^{-32}\,\mathrm{cm}^2$, and electric millicharge $q_{\nu_e} = +4.0\times 10^{-9}\,e$; and
(iii) flavor-conserving vector NSI with $\varepsilon_{ee}^{uV} = -0.05$.
Constructive enhancements from NSI ($\varepsilon_{ee}^{uV} = -0.05$), the light scalar $\phi$, the $U(1)_{B-L}$ mediator, and $|\mu_{\nu_e}^{\mathrm{eff}}|$ raise the low-energy rate above the SM band, whereas destructive interference in the universal $Z'$ model, negative $\langle r_{\nu_e}^2 \rangle$, and positive $q_{\nu_e}$ suppress the signal near threshold. Notably, whereas NSI and $\langle r_{\nu_e}^2 \rangle$ induce a nearly energy-independent rescaling of the SM spectrum, interactions exhibiting $1/T_{\mathcal{N}}$ or $1/(q^2 + m^2)$ scaling produce a steeper relative spectral distortion toward the lowest energy bins.

\begin{figure}[htbp]
    \centering
    \includegraphics[width=0.98\columnwidth]{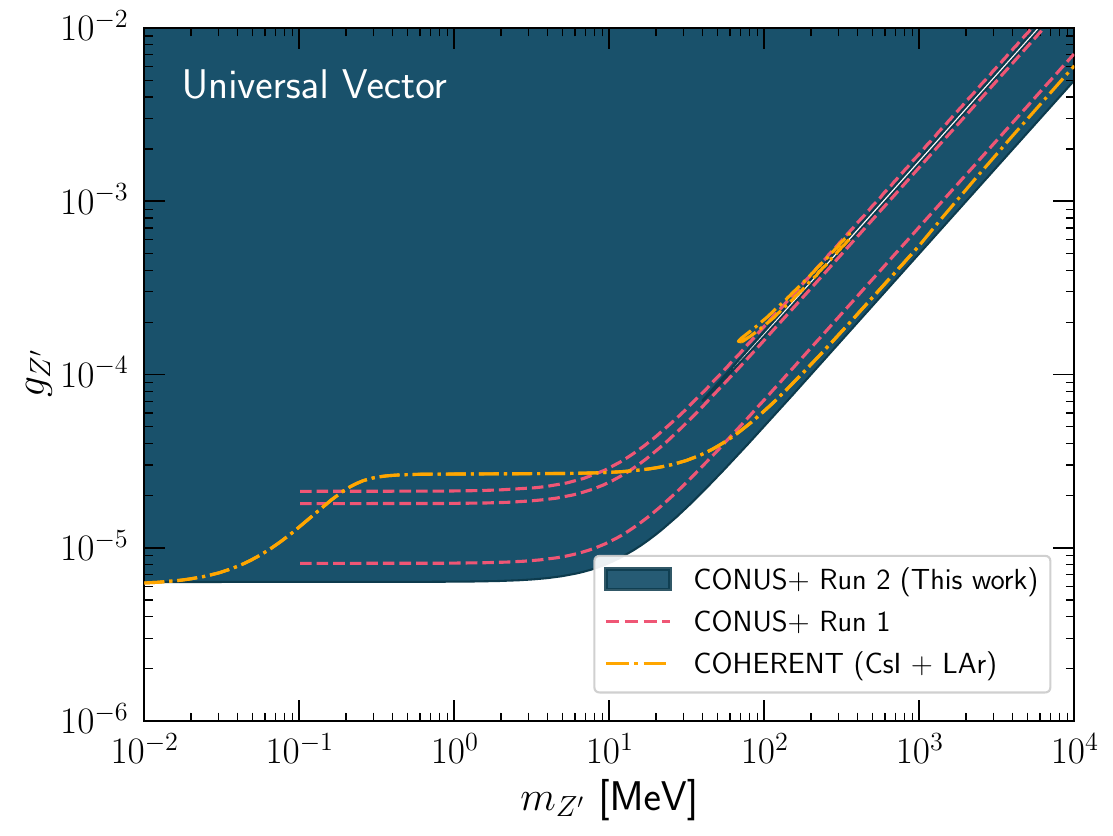}
    \caption{The $90\%$ C.L. exclusion region in the $(m_{Z'}, g_{Z'})$ plane for the universal light vector mediator model obtained from the CONUS+ Run~2 CE$\nu$NS data (shaded region), compared with the $90\%$ C.L. limits from CONUS+ Run~1~\cite{DeRomeri:2025csu} (dashed curve) and COHERENT (CsI+LAr)~\cite{DeRomeri:2022twg} (dash-dotted curve).}
    \label{fig:lv_universal}
\end{figure}

\begin{figure}[htbp]
    \centering
    \includegraphics[width=0.98\columnwidth]{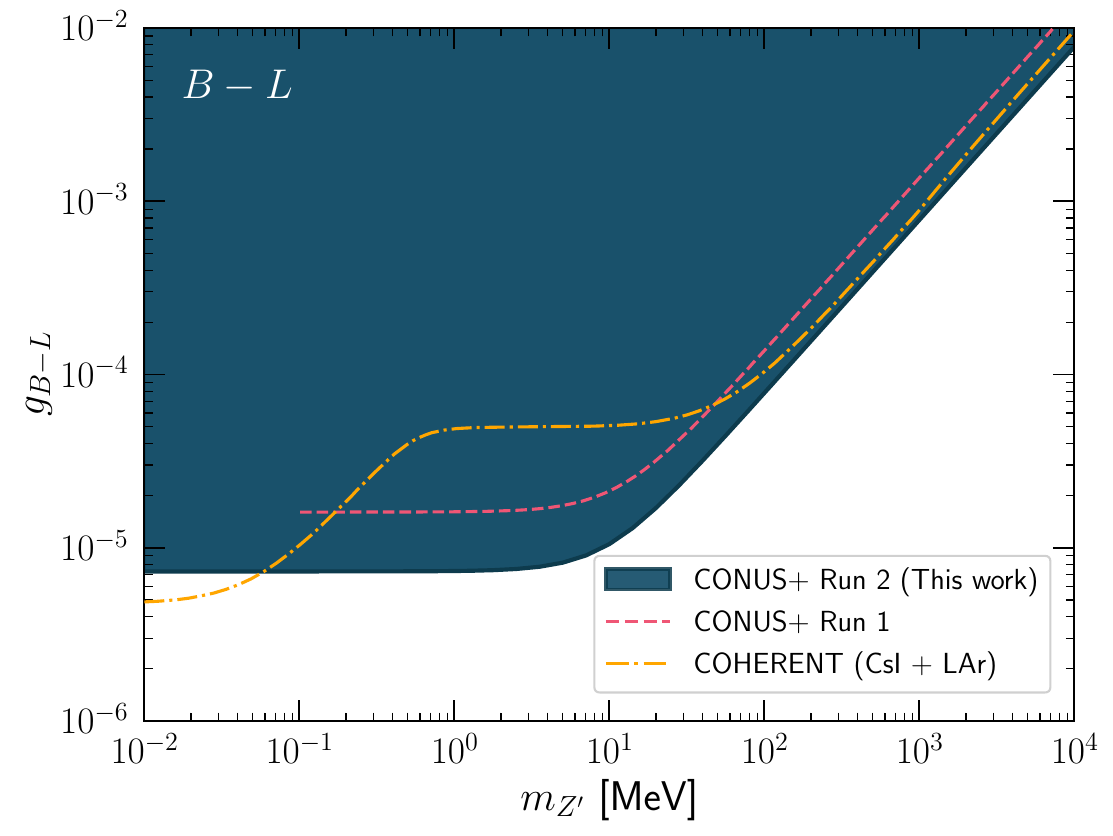}
    \caption{The $90\%$ C.L. exclusion region in the $(m_{Z'}, g_{B-L})$ plane for the anomaly-free $U(1)_{B-L}$ vector mediator model obtained from the CONUS+ Run~2 CE$\nu$NS data (shaded region), compared with the $90\%$ C.L. limits from CONUS+ Run~1~\cite{DeRomeri:2025csu} (dashed curve) and COHERENT (CsI+LAr)~\cite{DeRomeri:2022twg} (dash-dotted curve).}
    \label{fig:lv_bl}
\end{figure}

\subsection{Constraints on Light Mediators}
\label{sec:results_mediators}

\begin{figure}[ht]
    \centering
    \includegraphics[width=0.98\columnwidth]{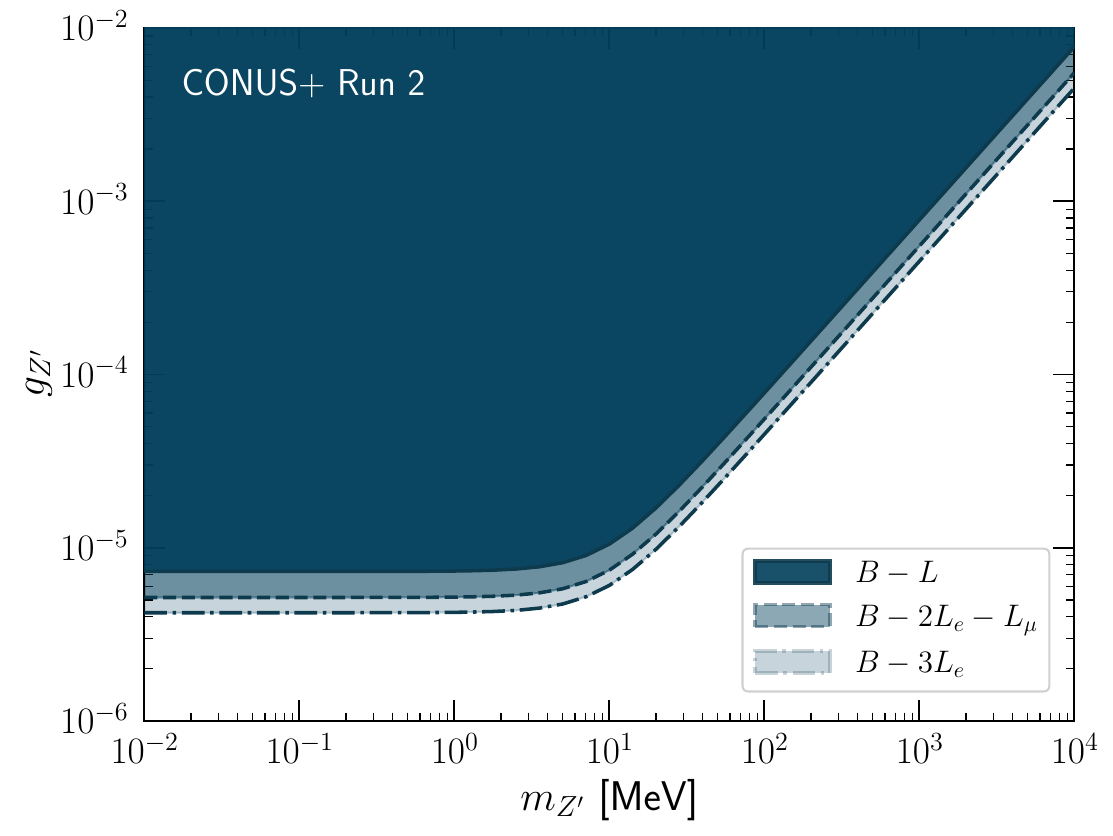}
    \caption{Comparison of the 90\% C.L. exclusion bounds in the $(m_{Z'}, g_{Z'})$ plane obtained from CONUS+ Run 2 CE$\nu$NS data for three anomaly-free $U(1)'$ gauge boson models: $B-L$ (solid curve), $B-2L_e-L_\mu$ (dashed curve), and $B-3L_e$ (dash-dotted curve).}
    \label{fig:lv_anomaly_free_comp}
\end{figure}

\begin{figure}[ht]
    \centering
    \includegraphics[width=0.98\columnwidth]{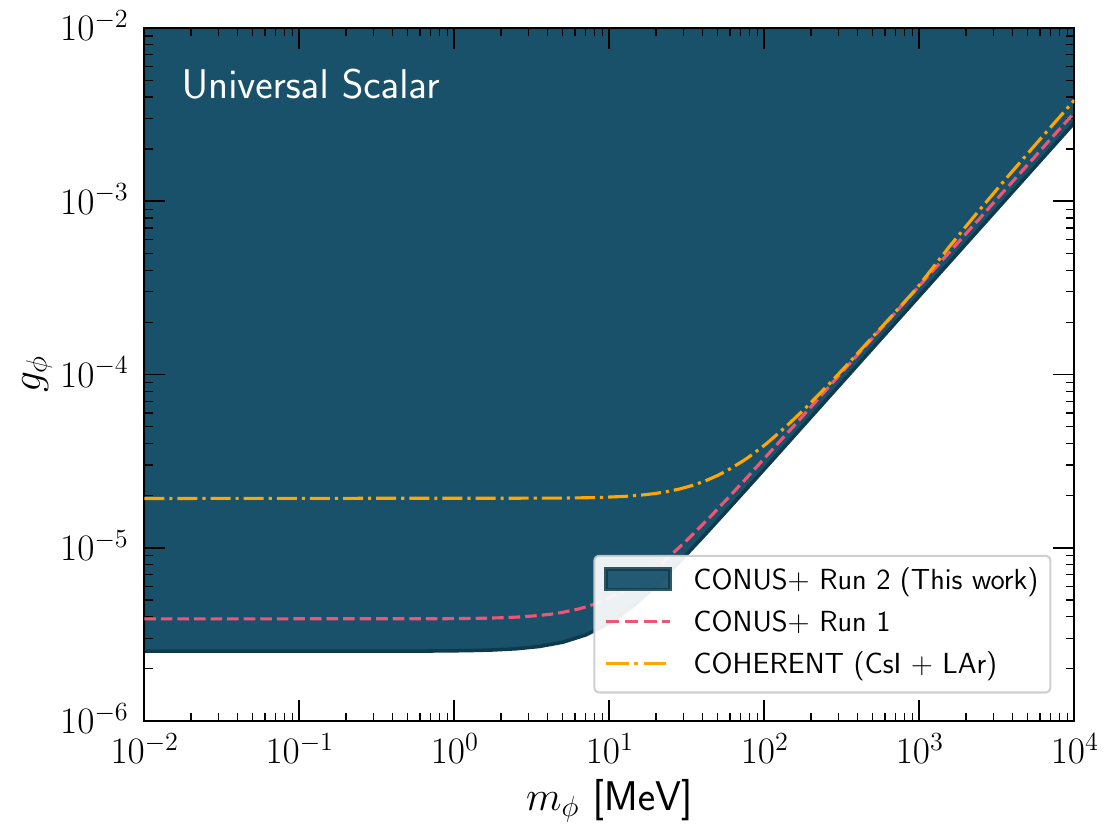}
    \caption{The $90\%$ C.L. exclusion region in the $(m_\phi, g_\phi)$ plane for the universal light scalar mediator model obtained from the CONUS+ Run~2 CE$\nu$NS data (shaded region), compared with the $90\%$ C.L. limits from CONUS+ Run~1~\cite{DeRomeri:2025csu} (dashed curve) and COHERENT (CsI+LAr)~\cite{DeRomeri:2022twg} (dash-dotted curve).}
    \label{fig:ls_scalar}
\end{figure}

We present the 90\% C.L. ($\Delta\chi^2 = 4.61$, 2 d.o.f.) exclusion bounds obtained from the CONUS+ Run 2 CE$\nu$NS dataset for light vector ($Z'$) and scalar ($\phi$) mediators in Figs.~\ref{fig:lv_universal}, \ref{fig:lv_bl},~\ref{fig:lv_anomaly_free_comp}, and \ref{fig:ls_scalar}. For comparison, the 90\% C.L. limits from CONUS+ Run 1~\cite{DeRomeri:2025csu} (dashed curve) and the combined COHERENT CsI + LAr dataset~\cite{DeRomeri:2022twg} (dash-dotted curve) are also shown. Across all scenarios, the exclusion contours display the two characteristic kinematic regimes set by the typical momentum transfer of reactor CE$\nu$NS in germanium, $|\vec{q}| = \sqrt{2 M_i T_{\mathcal{N}}} \sim \mathcal{O}(1\text{--}10)\;\mathrm{MeV}$: a mass-independent plateau in the light-mediator limit ($m_{Z',\phi} \ll |\vec{q}|$) and a linear scaling $g_{Z',\phi} \propto m_{Z',\phi}$ in the heavy-mediator contact-interaction regime ($m_{Z',\phi} \gg |\vec{q}|$).

In the universal light vector model (Fig.~\ref{fig:lv_universal}), where the $Z'$ boson couples equally to all first-generation quarks and electron neutrinos, the new-physics amplitude interferes destructively with the SM weak nuclear charge $Q_{\mathrm{SM},i}$. 
This destructive interference in principle gives rise to a narrow unconstrained degeneracy corridor in the $(m_{Z'}, g_{Z'})$ plane along $Q_{V,i}(T_{\mathcal{N}}) \approx -Q_{\mathrm{SM},i}$ (visible in the CONUS+ Run~1 and COHERENT curves), which for CONUS+ Run~2 is narrowed by the higher statistics to a width too thin to be visually resolved on the logarithmic scale of Fig.~\ref{fig:lv_universal}.
In the light-mediator plateau ($m_{Z'} \lesssim 1\;\mathrm{MeV}$), the CONUS+ Run~2 dataset sets a $90\%$ C.L. upper limit of
\begin{equation}
g_{Z'} \leq 6.34 \times 10^{-6}.
\end{equation}
Owing to the increased exposure and improved systematic control of Run~2, the excluded parameter space tightens visibly relative to CONUS+ Run~1 across the entire mediator mass range. Compared with the combined COHERENT (CsI+LAr) constraints, CONUS+ Run~2 provides substantially stronger sensitivity across the intermediate-to-low mass range ($0.1\;\mathrm{MeV} \lesssim m_{Z'} \lesssim 50\;\mathrm{MeV}$) due to the lower momentum transfer of reactor antineutrinos, while below $m_{Z'} \lesssim 0.3\;\mathrm{MeV}$ the COHERENT (CsI+LAr) curve bends downward owing to the inclusion of CsI E$\nu$ES events in Ref.~\cite{DeRomeri:2022twg}, approaching the CONUS+ Run~2 pure-CE$\nu$NS plateau near $m_{Z'} \sim 10^{-2}\;\mathrm{MeV}$. At higher mediator masses ($m_{Z'}\gtrsim 50$~MeV), the CONUS+ Run~2 and COHERENT (CsI+LAr) limits are comparable, and the two experiments are complementary.

For the $U(1)_{B-L}$ model (Fig.~\ref{fig:lv_bl}), the $Z'$ contribution adds constructively to $Q_{\mathrm{SM},i}$ in antineutrino-nucleus scattering, eliminating the destructive-interference degeneracy. In the light-mediator plateau ($m_{Z'} \lesssim 1\;\mathrm{MeV}$), CONUS+ Run~2 sets a $90\%$ C.L. upper bound of
\begin{equation}
g_{B-L} \leq 7.30 \times 10^{-6}.
\end{equation}
This represents a pronounced improvement over CONUS+ Run~1 throughout the full mass range. Relative to COHERENT (CsI+LAr), CONUS+ Run~2 significantly tightens the exclusion boundary over $0.1\;\mathrm{MeV} \lesssim m_{Z'} \lesssim 50\;\mathrm{MeV}$ and maintains slightly stronger sensitivity at higher mediator masses, whereas below $m_{Z'} \lesssim 0.06\;\mathrm{MeV}$ the COHERENT (CsI+LAr) limit extends below our pure-CE$\nu$NS plateau owing to the CsI E$\nu$ES contribution included in Ref.~\cite{DeRomeri:2022twg}.

In Fig.~\ref{fig:lv_anomaly_free_comp}, we compare the CONUS+ Run 2 exclusion bounds across three anomaly-free $U(1)'$ gauge symmetries by extending the analysis to the flavor-dependent models $U(1)_{B-2L_e-L_\mu}$ and $U(1)_{B-3L_e}$. In the light-mediator regime, these models yield the 90\% C.L. upper limits
\begin{align}
g_{B-2L_e-L_\mu} &\leq 5.16 \times 10^{-6}, \\
g_{B-3L_e} &\leq 4.21 \times 10^{-6}.
\end{align}
Because reactor CE$\nu$NS probes exclusively the $\bar{\nu}_e$ flux and all three anomaly-free symmetries share the same nuclear baryon charge ($Q'_N = A$), the effective coupling combination $g_{Z'}^2 Q'_{\nu_e} Q'_N$ scales strictly in the ratio $1 : 2 : 3$ across $B-L$, $B-2L_e-L_\mu$, and $B-3L_e$. Consequently, the resulting bounds follow the exact analytical hierarchy $g_{B-L} : g_{B-2L_e-L_\mu} : g_{B-3L_e} = 1 : 1/\sqrt{2} : 1/\sqrt{3}$ across the entire mediator mass range.

In the universal light scalar scenario (Fig.~\ref{fig:ls_scalar}), the scalar exchange does not interfere with the SM vector amplitude and benefits from the coherent scalar nuclear enhancement at low recoil energies. From CONUS+ Run 2, we derive a stringent 90\% C.L. upper limit in the light-mediator plateau of
\begin{equation}
g_{\phi} \leq 2.53 \times 10^{-6}.
\end{equation}
Compared to CONUS+ Run 1, Run 2 consistently tightens the exclusion boundary across both the low- and high-mass regimes. Moreover, CONUS+ Run 2 surpasses the combined COHERENT (CsI + LAr) sensitivity by a wide margin below $m_\phi \sim 10\;\mathrm{MeV}$, while at higher masses ($m_\phi \gtrsim 100\;\mathrm{MeV}$) the two limits become comparable, with CONUS+ Run 2 slightly stronger. This underscores the leading role of low-threshold reactor detectors among CE$\nu$NS probes of light scalar interactions.
\begin{table*}[hbt]
	\centering
	\caption{NSI bounds at $1\sigma$ C.L. ($\Delta\chi^2 = 1.00$) obtained from the CONUS+ Run 2 CE$\nu$NS dataset, compared with the CONUS+ Run 1~\cite{DeRomeri:2025csu} and combined COHERENT (CsI+LAr)~\cite{DeRomeri:2022twg} bounds.}
	\label{tab:nsi_bounds}
	\renewcommand{\arraystretch}{1.25}
	\begin{tabular*}{\textwidth}{@{\extracolsep{\fill}}cccc}
		\hline\hline
		\textbf{NSI} & \textbf{CONUS+ Run 2 (This work)} & \textbf{CONUS+ Run 1}~\cite{DeRomeri:2025csu} & \textbf{COHERENT (CsI+LAr)}~\cite{DeRomeri:2022twg} \\
		\hline
		$\varepsilon_{ee}^{uV}$   & $[-0.003,\, 0.021] \cup [0.352,\, 0.379]$ & $[-0.037,\, 0.026] \cup [0.348,\, 0.411]$ & $[-0.024,\, 0.045] \cup [0.34,\, 0.43]$ \\
		$\varepsilon_{ee}^{dV}$   & $[-0.004,\, 0.020] \cup [0.326,\, 0.350]$ & $[-0.034,\, 0.024] \cup [0.322,\, 0.380]$ & $[-0.027,\, 0.048] \cup [0.30,\, 0.39]$ \\
		$\varepsilon_{e\mu}^{uV}$ & $[-0.055,\, 0.055]$                       & $[-0.123,\, 0.123]$                       & $[-0.081,\, 0.081]$                    \\
		$\varepsilon_{e\mu}^{dV}$ & $[-0.051,\, 0.051]$                       & $[-0.114,\, 0.114]$                       & $[-0.071,\, 0.071]$                    \\
		$\varepsilon_{e\tau}^{uV}$& $[-0.055,\, 0.055]$                       & $[-0.123,\, 0.123]$                       & $[-0.13,\, 0.13]$                      \\
		$\varepsilon_{e\tau}^{dV}$& $[-0.051,\, 0.051]$                       & $[-0.114,\, 0.114]$                       & $[-0.12,\, 0.12]$                      \\
		\hline\hline
	\end{tabular*}
\end{table*}
%
\begin{figure*}[htb!]
	\centering
	\begin{minipage}[b]{0.47\textwidth}
		\centering
		\includegraphics[width=\textwidth]{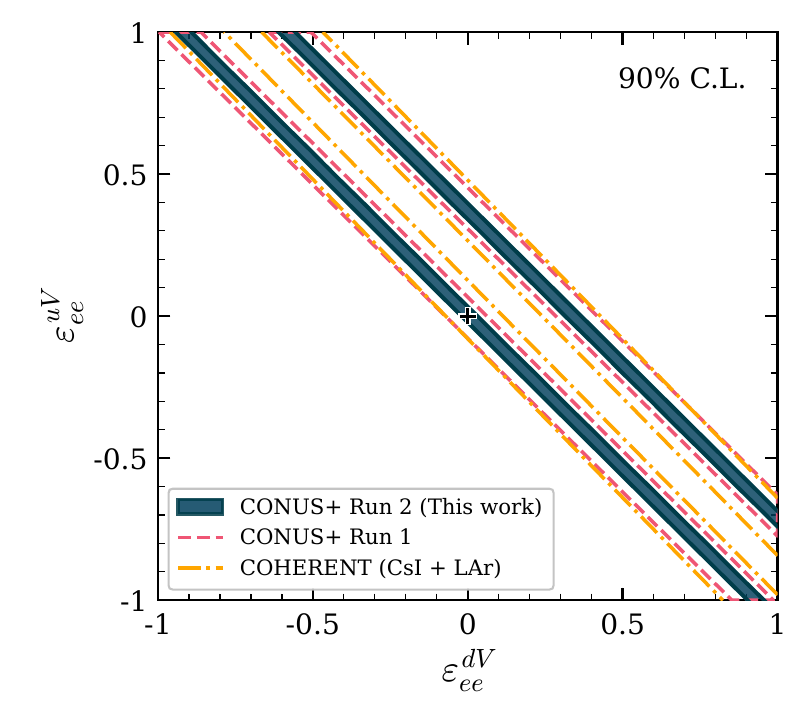}\\
		{\centering (a)}
	\end{minipage}
	\begin{minipage}[b]{0.47\textwidth}
		\centering
		\includegraphics[width=\textwidth]{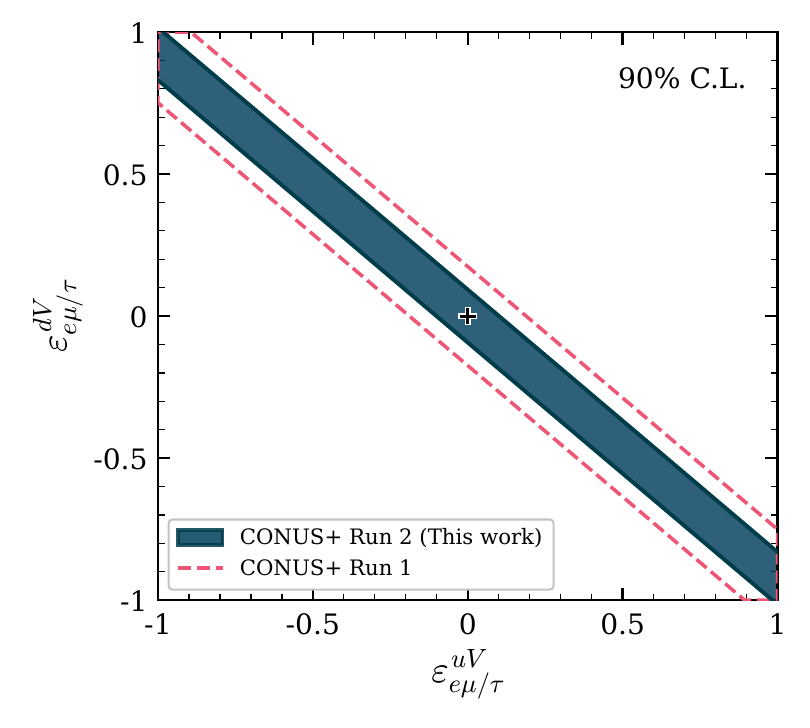}\\
		{\centering (b)}
	\end{minipage}\\[0.6em]
	\begin{minipage}[b]{0.47\textwidth}
		\centering
		\includegraphics[width=\textwidth]{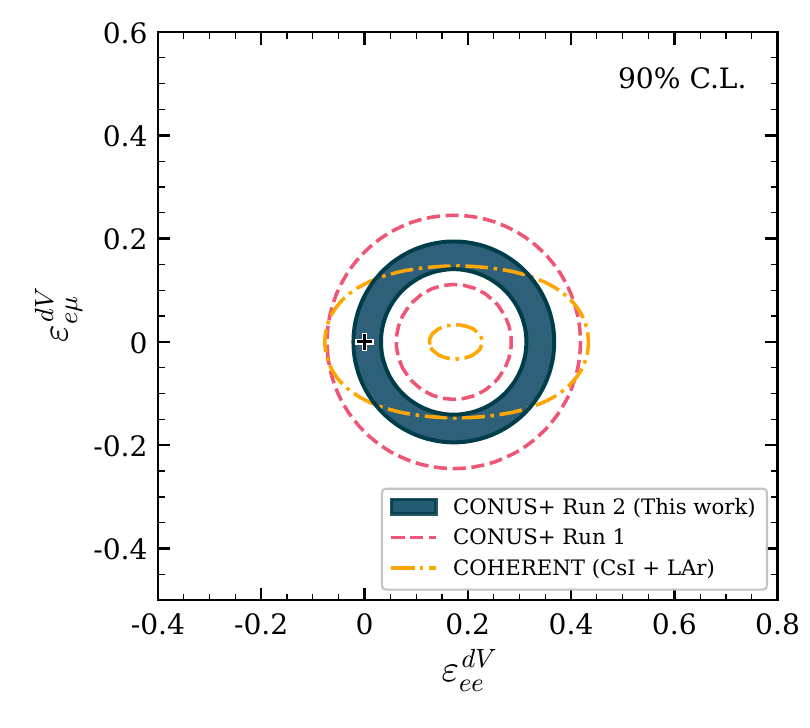}\\
		{\centering (c)}
	\end{minipage}
	\begin{minipage}[b]{0.47\textwidth}
		\centering
		\includegraphics[width=\textwidth]{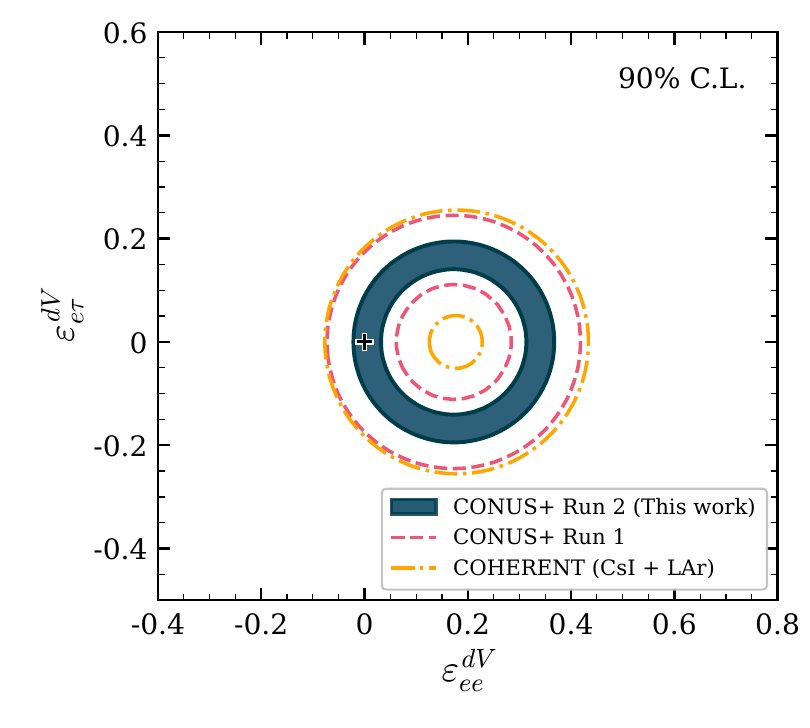}\\
		{\centering (d)}
	\end{minipage}
	\caption{Allowed $90\%$ C.L. regions (2~d.o.f.) in the two-dimensional NSI parameter planes from our CONUS+ Run~2 analysis (shaded teal regions), compared with existing CE$\nu$NS bounds from CONUS+ Run~1 (dashed red)~\cite{DeRomeri:2025csu} and COHERENT CsI+LAr (dot-dashed yellow)~\cite{DeRomeri:2022twg}, shown where available: (a)~$(\varepsilon_{ee}^{dV}, \varepsilon_{ee}^{uV})$, (b)~$(\varepsilon_{e\mu/\tau}^{uV}, \varepsilon_{e\mu/\tau}^{dV})$, (c)~$(\varepsilon_{ee}^{dV}, \varepsilon_{e\mu}^{dV})$, and (d)~$(\varepsilon_{ee}^{dV}, \varepsilon_{e\tau}^{dV})$. The black cross marks the SM point $(0, 0)$.}
	\label{fig:nsi_2d_comparison}
\end{figure*}

\subsection{Constraints on Non-standard Neutrino Interactions}
\label{sec:results_nsi}
We first summarize the one-dimensional $1\sigma$ C.L. ($\Delta\chi^2 = 1.00$) bounds on the flavor-conserving ($\varepsilon_{ee}^{uV}$, $\varepsilon_{ee}^{dV}$) and flavor-changing ($\varepsilon_{e\mu}^{uV,dV}$, $\varepsilon_{e\tau}^{uV,dV}$) NSI parameters derived from the CONUS+ Run~2 CE$\nu$NS dataset in Table~\ref{tab:nsi_bounds}, alongside a comparison with the CONUS+ Run~1~\cite{DeRomeri:2025csu} (see also Ref.~\cite{Benavides:2026conus} for a recent independent Run~1 NSI analysis) and combined COHERENT (CsI+LAr)~\cite{DeRomeri:2022twg} limits. In each one-parameter scan, only a single NSI coupling is varied while all other couplings are fixed to zero.

For the flavor-conserving couplings $\varepsilon_{ee}^{uV}$ and $\varepsilon_{ee}^{dV}$, the quadratic dependence of the CE$\nu$NS cross section on the modified weak nuclear charge $Q_{\mathrm{NSI},i}$ gives rise to a twofold degeneracy ($Q_{\mathrm{NSI},i} \approx \pm Q_{\mathrm{SM},i}$),
splitting the $1\sigma$ allowed regions into two disjoint intervals: $[-0.003,\, 0.021] \cup [0.352,\, 0.379]$ for $\varepsilon_{ee}^{uV}$ and $[-0.004,\, 0.020] \cup [0.326,\, 0.350]$ for $\varepsilon_{ee}^{dV}$. For the flavor-changing couplings $\varepsilon_{e\mu}^{qV}$ and $\varepsilon_{e\tau}^{qV}$, the absence of interference with the SM amplitude yields symmetric bounds around zero, which are identical for the $\mu$ and $\tau$ channels in a pure $\bar{\nu}_e$ reactor flux: $|\varepsilon_{e\mu}^{uV}|, |\varepsilon_{e\tau}^{uV}| \leq 0.055$ and $|\varepsilon_{e\mu}^{dV}|, |\varepsilon_{e\tau}^{dV}| \leq 0.051$ at $1\sigma$ C.L. Across all six one-dimensional parameters, CONUS+ Run~2 shrinks the allowed intervals by more than a factor of two compared to CONUS+ Run~1 and places tighter constraints than the combined COHERENT (CsI+LAr) dataset.

\begin{figure*}[htbp]
	\centering
	\includegraphics[width=\textwidth]{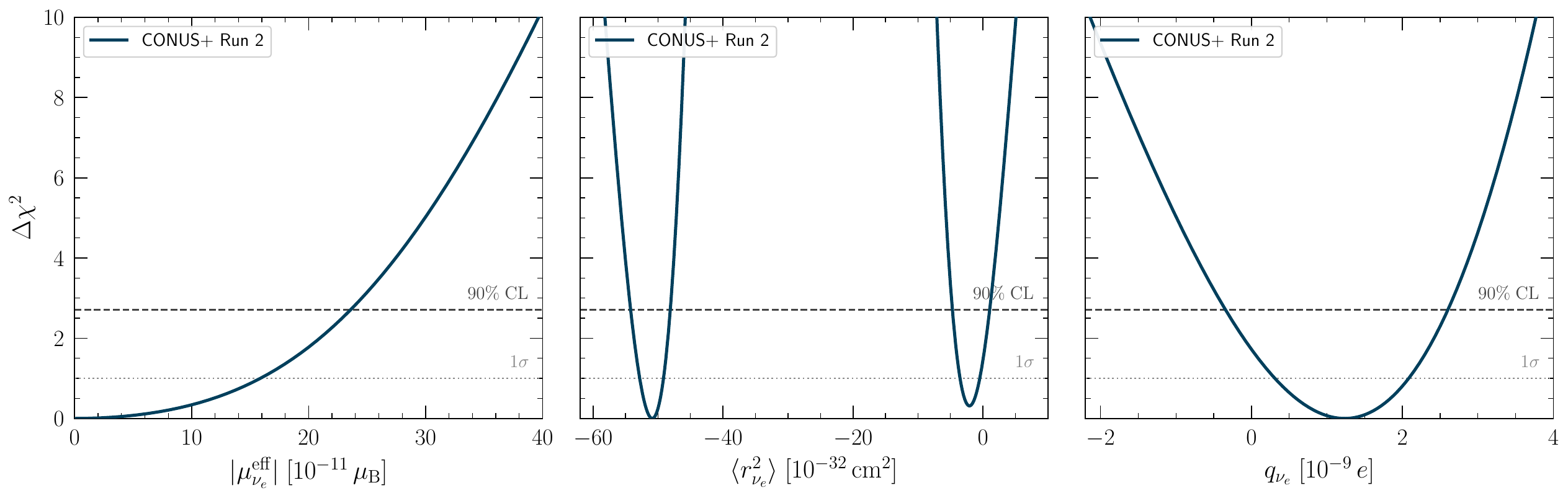}
	\caption{$\Delta\chi^2$ profiles obtained from the CONUS+ Run 2 CE$\nu$NS dataset for the electron neutrino electromagnetic properties: the effective magnetic moment $|\mu_{\nu_e}^{\mathrm{eff}}|$ (left), the charge radius $\langle r_{\nu_e}^2 \rangle$ (middle), and the electric millicharge $q_{\nu_e}$ (right). 
	}
	\label{fig:em_all_profiles}
\end{figure*}
To capture correlations and degeneracies among NSI couplings, we also perform two-parameter analyses at $90\%$ C.L. ($\Delta\chi^2 = 4.61$, 2~d.o.f.), presented in Fig.~\ref{fig:nsi_2d_comparison} for four representative parameter planes: $(\varepsilon_{ee}^{dV}, \varepsilon_{ee}^{uV})$, $(\varepsilon_{e\mu/\tau}^{uV}, \varepsilon_{e\mu/\tau}^{dV})$, $(\varepsilon_{ee}^{dV}, \varepsilon_{e\mu}^{dV})$, and $(\varepsilon_{ee}^{dV}, \varepsilon_{e\tau}^{dV})$, compared with CONUS+ Run~1 and COHERENT (CsI+LAr) where available.

In the flavor-conserving plane $(\varepsilon_{ee}^{dV}, \varepsilon_{ee}^{uV})$ [Fig.~\ref{fig:nsi_2d_comparison}(a)], the allowed region forms two parallel diagonal stripes with slope determined by the germanium proton-to-neutron ratio, $-(Z+2N)/(2Z+N)$. 
One band passes through the SM point $(0,0)$, while the second corresponds to the destructive-interference degeneracy $Q_{\mathrm{NSI},i} \approx -Q_{\mathrm{SM},i}$. CONUS+ Run~2 narrows both bands significantly relative to CONUS+ Run~1 and COHERENT, although a single nuclear target cannot break the linear combination degeneracy between $u$- and $d$-quark couplings without combining targets of different $N/Z$ ratios.

In the flavor-changing plane $(\varepsilon_{e\mu/\tau}^{uV}, \varepsilon_{e\mu/\tau}^{dV})$ [Fig.~\ref{fig:nsi_2d_comparison}(b)], the absence of SM interference collapses the allowed parameter space into a single diagonal band centered on the SM point $(0,0)$ with the reciprocal target-dependent slope $-(2Z+N)/(Z+2N)$. Because reactor antineutrinos are purely $\bar{\nu}_e$, the resulting contours are identical for the $e\mu$ and $e\tau$ sectors, with CONUS+ Run~2 visibly tightening the band width compared to CONUS+ Run~1.

Finally, when a flavor-conserving and a flavor-changing coupling are varied simultaneously---$(\varepsilon_{ee}^{dV}, \varepsilon_{e\mu}^{dV})$ in Fig.~\ref{fig:nsi_2d_comparison}(c) and $(\varepsilon_{ee}^{dV}, \varepsilon_{e\tau}^{dV})$ in Fig.~\ref{fig:nsi_2d_comparison}(d)---the quadratic sum of the coherent and flavor-changing amplitudes produces an annular (ring-shaped) allowed region passing through the SM point $(0,0)$. In both planes, CONUS+ Run~2 reduces the radial thickness of the allowed ring considerably compared to CONUS+ Run~1. In the $(\varepsilon_{ee}^{dV}, \varepsilon_{e\mu}^{dV})$ plane, the COHERENT (CsI+LAr) contour exhibits a slightly elliptical deformation because spallation-source neutrinos contain a substantial $\nu_\mu/\bar{\nu}_\mu$ component, providing enhanced sensitivity along the $\varepsilon_{e\mu}^{dV}$ axis, whereas CONUS+ Run~2 provides tighter control along the $\varepsilon_{ee}^{dV}$ direction. In contrast, in the $(\varepsilon_{ee}^{dV}, \varepsilon_{e\tau}^{dV})$ plane---where neither experiment has a $\nu_\tau$ flux---both contours remain annular, and CONUS+ Run~2 yields a noticeably narrower allowed ring than both CONUS+ Run~1 and COHERENT (CsI+LAr).

\begin{table*}[htbp]
\centering
\caption{Comparison of $90\%$ C.L. bounds on the electron neutrino ($\nu_e$) effective magnetic moment, electric millicharge, and charge radius derived from the CONUS+ Run~2 pure CE$\nu$NS dataset with existing limits from reactor, spallation-source, solar, dark matter direct detection, and accelerator experiments.}
\label{tab:em_nu_e_comparison}
\renewcommand{\arraystretch}{1.24}
\begin{tabular*}{\textwidth}{@{\extracolsep{\fill}}lccc}
\hline\hline
\textbf{Experiment} & $|\mu_{\nu_e}^{\mathrm{eff}}|\;(10^{-11}\,\mu_{\mathrm{B}})$ & $q_{\nu_e}\;(10^{-11}\,e)$ & $\langle r_{\nu_e}^2 \rangle\;(10^{-32}\,\mathrm{cm}^2)$ \\
\hline
\textbf{CONUS+ Run~2 (This work)} & $\mathbf{\leq 23.58}$ & $\mathbf{[-34.3,\, 260.2]}$ & $\mathbf{[-54.21,\, -48.11] \cup [-4.70,\, 1.03]}$ \\
CONUS+ Run~1 (CE$\nu$NS)~\cite{AtzoriCorona:2025ygn} & $\leq 56.0$ & --- & $[-76,\, -57] \cup [-8,\, 11]$ \\
CONUS+ Run 1 (CE$\nu$NS)~\cite{DeRomeri:2025csu}     & $\leq 47.8$ & --- & --- \\
TEXONO (CE$\nu$NS)~\cite{Ahmed:2026ge} & $\leq 59.6$ & $[-627,\, 3936]$ & $[-62.18,\, 9.37]$ \\
\hline
COHERENT Ge-mini (CE$\nu$NS+E$\nu$ES)~\cite{Ahmed:2026ge} & $\leq 323.5$ & $[-70.9,\, 71.7]$ & $[-60.17,\, -30.65] \cup [-22.27,\, 7.20]$ \\
Dresden-II (CE$\nu$NS+E$\nu$ES)~\cite{AtzoriCorona:2022qrf} & $\leq 21.3$ & $[-0.93,\, 0.95]$ & $[-56.7,\, -40.8] \cup [-11.6,\, 4.0]$ \\
COHERENT (CsI+LAr, CE$\nu$NS+E$\nu$ES)~\cite{AtzoriCorona:2022qrf} & $\leq 420.0$ & $[-50.0,\, 50.0]$ & $[-69.3,\, -49.2] \cup [-6.9,\, 14.4]$ \\
\hline
PandaX-4T~\cite{Demirci:2025em} & $\leq 1.87$ & $[-0.046,\, 0.063]$ & $[-114.71,\, 32.13]$ \\
XENONnT~\cite{Demirci:2025em} & $\leq 0.98$ & $[-0.018,\, 0.051]$ & $[-95.15,\, 13.53]$ \\
LZ~\cite{AtzoriCorona:2022jeb,Giunti:2023yha} & $\leq 1.5$ & $[-0.021,\, 0.020]$ & $[-110.4,\, 26.4]$ \\
TEXONO (E$\nu$ES)~\cite{TEXONO:2006xds,Chen:2014dsa,TEXONO:2009knm} & $\leq 7.4$ & $[-0.21,\, 0.21]$ & $[-4.2,\, 6.6]$ \\
GEMMA~\cite{Beda:2012zz} & $\leq 2.9$ & --- & --- \\
Borexino~\cite{Borexino:2017fbd,Coloma:2022umy} & $\leq 3.9$ & --- & --- \\
LSND~\cite{LSND:2001fbw} & --- & --- & $[-5.94,\, 8.28]$ \\
\hline\hline
\end{tabular*}
\end{table*}

\subsection{Constraints on Neutrino Electromagnetic Properties}
\label{sec:results_em}

In Fig.~\ref{fig:em_all_profiles}, we display the one-dimensional $\Delta\chi^2$ profiles obtained from the CONUS+ Run~2 pure CE$\nu$NS dataset for the electron neutrino effective magnetic moment $|\mu_{\nu_e}^{\mathrm{eff}}|$, charge radius $\langle r_{\nu_e}^2 \rangle$, and electric millicharge $q_{\nu_e}$. The resulting $90\%$ C.L. ($\Delta\chi^2 = 2.71$, 1~d.o.f.) limits are summarized in Table~\ref{tab:em_nu_e_comparison}, together with existing reactor, spallation-source, solar, direct detection, and accelerator bounds and their corresponding references.

For the effective electron neutrino magnetic moment (left panel of Fig.~\ref{fig:em_all_profiles}), the CONUS+ Run~2 profile favors a vanishing magnetic moment ($|\mu_{\nu_e}^{\mathrm{eff}}| = 0$), consistent with the SM expectation, setting a $90\%$ C.L. pure-CE$\nu$NS upper limit of $|\mu_{\nu_e}^{\mathrm{eff}}| \leq 23.58 \times 10^{-11}\,\mu_{\mathrm{B}}$. 
Compared on an equal pure-CE$\nu$NS footing with the two independent CONUS+ Run~1 analyses in Table~\ref{tab:em_nu_e_comparison}, which differ slightly owing to their respective statistical and systematic prescriptions, our Run~2 result tightens the nuclear-recoil magnetic moment bound by approximately a factor of two. 
Owing to the $1/T_{\mathcal{N}}$ enhancement at low reactor recoil energies, our limit also improves by more than a factor of two upon the CE$\nu$NS-only sensitivity of TEXONO and surpasses spallation-source bounds---such as COHERENT Ge-mini and COHERENT (CsI+LAr)---by more than an order of magnitude, even when E$\nu$ES contributions are included in those analyses. When the E$\nu$ES channel is combined with CE$\nu$NS, reactor analyses such as Dresden-II, as well as combined CONUS+ Run~1 and TEXONO studies, achieve tighter constraints, while dedicated low-background solar and reactor E$\nu$ES experiments---led by XENONnT, PandaX-4T, LZ, GEMMA, Borexino, and TEXONO---currently provide the strongest laboratory bounds.

For the electron neutrino charge radius (middle panel of Fig.~\ref{fig:em_all_profiles}), the quadratic dependence of the CE$\nu$NS cross section on the modified weak nuclear charge yields two nearly degenerate minima, separating the $90\%$ C.L. allowed region into two narrow, disjoint intervals, $\langle r_{\nu_e}^2 \rangle \in [-54.21,\, -48.11] \cup [-4.70,\, 1.03] \times 10^{-32}\,\mathrm{cm}^2$. The SM prediction, $\langle r_{\nu_e}^2 \rangle_{\mathrm{SM}} = -0.83 \times 10^{-32}\,\mathrm{cm}^2$, lies comfortably within the SM-compatible branch. Because the E$\nu$ES contribution to $\langle r_{\nu_e}^2 \rangle$ is negligible compared to the coherently enhanced CE$\nu$NS rate in reactor CE$\nu$NS setups, pure CE$\nu$NS datasets are directly competitive with combined CE$\nu$NS+E$\nu$ES determinations.
Thanks to its higher statistics and improved systematic control, CONUS+ Run~2 decisively excludes the intermediate region between the two minima at $90\%$ C.L.---in contrast to the single broad interval obtained for TEXONO---and yields markedly narrower allowed branches than CONUS+ Run~1, COHERENT Ge-mini, Dresden-II, and COHERENT (CsI+LAr). Moreover, the CONUS+ Run~2 charge radius intervals are substantially tighter than the flavor-dependent solar E$\nu$ES bounds from PandaX-4T, XENONnT, and LZ. Notably, the width of the SM-compatible branch is even narrower than those of dedicated E$\nu$ES measurements such as LSND and TEXONO (E$\nu$ES), although E$\nu$ES data remain essential to break the degeneracy with the negative branch.

For the electron neutrino electric millicharge (right panel of Fig.~\ref{fig:em_all_profiles}), the CONUS+ Run~2 pure CE$\nu$NS analysis exhibits a mild ($\sim 1.3\sigma$) preference for a small positive charge due to destructive interference with the SM amplitude at low recoil energies, while remaining consistent with electrical neutrality ($q_{\nu_e} = 0$) at $90\%$ C.L., yielding the allowed interval $q_{\nu_e} \in [-0.343,\, 2.602] \times 10^{-9}\,e$.
On an equal CE$\nu$NS-only footing, CONUS+ Run~2 improves upon the TEXONO CE$\nu$NS-only sensitivity by more than an order of magnitude; notably, our pure-CE$\nu$NS reactor bound is within a factor of a few of the spallation-source CE$\nu$NS+E$\nu$ES limits from COHERENT Ge-mini and COHERENT (CsI+LAr).
Nevertheless, because the millicharge cross section scales as $1/|\vec{q}|^4$, incorporating the E$\nu$ES channel in reactor setups enhances sensitivity by roughly two to three orders of magnitude relative to nuclear recoils alone. Accordingly, combined CE$\nu$NS+E$\nu$ES reactor analyses---such as Dresden-II and combined CONUS+ Run~1 and TEXONO fits---alongside low-energy solar and reactor E$\nu$ES experiments (XENONnT, PandaX-4T, LZ, and TEXONO) yield considerably tighter millicharge constraints, whereas our CONUS+ Run~2 bound provides a clean nuclear-recoil benchmark independent of low-energy atomic-ionization modeling uncertainties in the germanium E$\nu$ES sector.

\section{Conclusions}
\label{sec:conc}

In this work, we have performed a combined binned Poisson likelihood analysis of the precision CONUS+ Run~2 CE$\nu$NS dataset ($1296\;\mathrm{kg\cdot days}$ of reactor-on exposure across three germanium detectors) to probe physics beyond the SM at low momentum transfer. Specifically, we derived updated constraints on light vector and scalar mediators, NSIs, and electron neutrino electromagnetic properties, and compared them with CONUS+ Run~1 as well as complementary spallation-source, solar, direct detection, and accelerator measurements.

For light mediators in the low-mass plateau ($m_{Z',\phi} \lesssim 1\;\mathrm{MeV}$), CONUS+ Run~2 sets $90\%$ C.L. upper bounds of $g_{Z'} \leq 6.34 \times 10^{-6}$ for the universal vector mediator and $g_\phi \leq 2.53 \times 10^{-6}$ for the universal scalar mediator, while the anomaly-free gauge extensions yield $g_{B-L} \leq 7.30 \times 10^{-6}$, $g_{B-2L_e-L_\mu} \leq 5.16 \times 10^{-6}$, and $g_{B-3L_e} \leq 4.21 \times 10^{-6}$, obeying the exact $1 : 1/\sqrt{2} : 1/\sqrt{3}$ scaling. In the NSI sector, our one-dimensional $1\sigma$ analysis shrinks the allowed intervals by more than a factor of two relative to CONUS+ Run~1 and surpasses combined COHERENT (CsI+LAr) limits across all flavor-conserving and flavor-changing parameters, while the two-dimensional $90\%$ C.L. contours substantially narrow the correlated diagonal bands and annular regions. For the electron neutrino electromagnetic properties, our pure-CE$\nu$NS analysis yields $90\%$ C.L. bounds of $|\mu_{\nu_e}^{\mathrm{eff}}| \leq 23.58 \times 10^{-11}\,\mu_{\mathrm{B}}$, $\langle r_{\nu_e}^2 \rangle \in [-54.21,\, -48.11] \cup [-4.70,\, 1.03] \times 10^{-32}\,\mathrm{cm}^2$, and $q_{\nu_e} \in [-0.343,\, 2.602] \times 10^{-9}\,e$. Notably, CONUS+ Run~2 decisively excludes the intermediate charge-radius region between the two nearly degenerate minima and establishes a clean nuclear-recoil benchmark for $|\mu_{\nu_e}^{\mathrm{eff}}|$ and $q_{\nu_e}$ independent of low-energy atomic-ionization modeling uncertainties.

Overall, these results show that the fourfold exposure increase and improved systematic control of CONUS+ Run~2 establish reactor-based CE$\nu$NS as a competitive probe of new neutrino interactions, complementing spallation-source measurements on different nuclear targets as well as solar and reactor E$\nu$ES experiments. Because the present bounds are already affected by the $11.6\%$ signal-prediction uncertainty---dominated by the threshold energy scale and Lindhard quenching factor---together with residual background-shape uncertainties, future progress will require not only larger exposures but also improved low-energy calibration and background modeling. Combining such refinements with dedicated E$\nu$ES channels and multi-target CE$\nu$NS datasets will be essential to break remaining parameter degeneracies and further tighten constraints on physics beyond the SM.

\section*{Acknowledgments}
This work was supported by the Scientific and Technological Research Council of Türkiye (TUBITAK) under the project numbers 124F416 and 126F009. 
\bibliographystyle{unsrt}

\end{document}